\documentclass[aps,superscriptaddress,twocolumn,pre,longbibliography]{revtex4-1}

\usepackage{amsmath,graphicx,color,epsfig,latexsym,bm,ulem,dutchcal,subfigure,float,tikz,eucal,mathpazo,times,braket}
\usepackage{amssymb}
\usepackage{bbm}
\usepackage[colorlinks=true,linkcolor=blue,urlcolor=blue,citecolor=blue,anchorcolor=red]{hyperref}

\usetikzlibrary{positioning}

\begin{document}

\title{Engineering a Quantum Thermal Diode with Floquet Driving}

\author{Aqsa Rehman}
\affiliation{Department of Physics, Lahore University of Management Sciences, Lahore 54792, Pakistan}

\author{M. Tahir Naseem}
\email{mnaseem16@ku.edu.tr}
\affiliation{Faculty of Basic Sciences, Ghulam Ishaq Khan Institute of Engineering Sciences and Technology, Topi 23640, District Swabi, Khyber Pakhtunkhwa, Pakistan}
\affiliation{National Center for Quantum Computing, University of Engineering and Technology Narowal, 10 km Muridke Road, Near Adda Siraj, Narowal, Pakistan}

\author{Adam Zaman Chaudhry}
\email{adam.zaman@lums.edu.pk}
\affiliation{Department of Physics, Lahore University of Management Sciences, Lahore 54792, Pakistan}

\date{\today}

\begin{abstract}
Controlling heat flow in small quantum systems is a central goal of quantum thermodynamics and nanoscale transport.
A key challenge is to achieve strong thermal rectification without suppressing the transmitted heat current, a tradeoff that often arises in static diode configurations.
We establish a Floquet-control mechanism in a minimal quantum thermal diode formed by two longitudinally modulated Ising-coupled qubits, each coupled to an independent thermal reservoir. From a microscopic system--bath model, we derive a Floquet--LGKS master equation that resolves the drive-assisted transition channels. The resonant undriven device with left--right symmetric bath couplings serves as the reciprocal benchmark, while static detuning provides a rectifying reference with reduced heat current.
Single-side driving breaks this reciprocal structure by creating a Floquet-dressed contact opposite a purely thermal contact. For this configuration, we obtain a compact steady-state current formula and an exact blocking condition for suppressing one bias direction while retaining finite transport in the opposite direction. In the weak sinusoidal-drive regime, rectification begins quadratically in the modulation amplitude. Dual-side driving adds a Floquet pumping contribution to the interaction-mediated current, so complete blocking requires cancellation of both contributions. These results establish contact-selective Floquet dressing as a design principle for controllable heat-flow asymmetry in minimal quantum thermal devices.
\end{abstract}

\maketitle


\section{Introduction}

Controlling heat flow at the nanoscale is a central problem in nonequilibrium physics, phononics, and quantum thermodynamics \cite{RevModPhys.84.1045,RevModPhys.93.041001,10.1116/5.0083192, Naseem2024HeatCurrents}. Beyond its fundamental relevance for understanding energy exchange in small systems, this problem also motivates the development of thermal devices that perform functions analogous to electronic components \cite{RevModPhys.83.131,Pekola2015,Vinjanampathy01102016,Goold_2016,Millen_2016}. Thermal diodes, logic elements, and transistors are prominent examples, providing basic building blocks for directional heat-flow control and thermal information processing \cite{RevModPhys.84.1045,PhysRevLett.99.177208,10.1063/1.2191730}. Progress in quantum heat transport has further extended these ideas to microscopic and mesoscopic platforms, where fluctuations, coherence, and reservoir-induced transitions can play an active role \cite{RevModPhys.81.1665,RevModPhys.83.771,RevModPhys.92.041002,annurev-conmatphys-062910-140506,MAHLER200553,Lebon2008}.

Within this landscape, thermal rectification provides one of the most direct manifestations of directional heat-flow control \cite{PhysRevLett.88.094302,PhysRevLett.95.104302,science.1132898,ROBERTS2011648}. A thermal diode supports unequal heat currents when the temperature bias is reversed, and this asymmetric response typically requires broken left-right symmetry together with a nonlinear transport mechanism \cite{PhysRevLett.93.184301,PhysRevLett.94.034301,PhysRevB.79.144306,PhysRevE.84.061135}. The concept was first developed in classical phononic systems \cite{PhysRevLett.88.094302,PhysRevLett.93.184301,PhysRevLett.95.104302,science.1132898} and was later extended to quantum platforms, including spin-boson models \cite{PhysRevLett.94.034301}, nonlinear quantum circuits \cite{PhysRevB.79.144306}, nonlinear resonators \cite{PhysRevB.103.155434, Naseem_2026}, minimal oscillator setups \cite{PhysRevE.103.012134}, nonlinear quantum chains \cite{Motz_2018}, and mesoscopic or superconducting architectures \cite{Senior2020,PhysRevApplied.11.044073}. This broader effort has also motivated thermal logic and transistor-like functionalities, where negative differential thermal resistance and heat-current amplification enable active control of energy transport \cite{10.1063/1.2191730,PhysRevLett.99.177208,PhysRevLett.116.200601,PhysRevB.101.184510,PhysRevResearch.2.033285,PhysRevApplied.16.034026}.

Few-body quantum systems provide especially useful benchmarks because they can display diode behavior without requiring extended many-body structures, while still allowing the relevant level structure and transition rates to be analyzed explicitly \cite{PhysRevLett.94.034301,PhysRevE.95.022128}. Interacting spins and qubits are particularly natural in this regard: their internal coupling creates state-dependent transition frequencies, and directional response can be induced through detuning, asymmetric reservoir couplings, or spectral mismatch \cite{PhysRevE.95.022128,PhysRevE.99.042121,PhysRevE.104.054137}. This mechanism has been explored in coupled-spin thermal diodes \cite{PhysRevE.95.022128,PhysRevE.104.054137,PhysRevE.107.064125}, quantum-optical two-atom configurations \cite{PhysRevE.99.042121}, superconducting artificial atoms \cite{Senior2020}, coupled-qubit photonic structures \cite{PhysRevApplied.15.054050}, and related thermal-machine settings \cite{Santiago-García_2025}. Extensions to spin chains and other interacting platforms further show how internal interactions and spectral structure can support directed heat flow beyond the minimal two-body limit \cite{PhysRevE.90.042142,PhysRevE.99.032116,PhysRevE.99.032136,PhysRevE.102.062146,Pereira_2019}. These studies make the static interacting-qubit diode a useful reference point for assessing what additional control can be gained from time-periodic driving.

Although static few-body devices provide useful benchmarks, they offer limited flexibility once the transition frequencies, coupling asymmetries, and reservoir spectra are fixed. Dynamical modulation provides an additional control knob because it can reshape the system--bath transition structure rather than merely tune static parameters \cite{PhysRevE.99.032126}. In a Floquet description, this reshaping appears through drive-assisted sidebands, which redistribute transition weights among multiple energy-exchange channels and can generate transport pathways absent in the undriven system. When the modulation is applied asymmetrically, these sideband-resolved channels can make the two contacts respond differently under temperature reversal, providing a mechanism for controllable rectification. Related Floquet-based studies have shown that time-periodic modulation can also enable broader thermal functionalities, including transistor-like behavior at fixed bath temperatures \cite{Gupt2022}. Floquet engineering therefore offers a natural route for extending minimal quantum thermal diodes beyond static rectification mechanisms.

In this work, we develop and analyze a minimal driven thermal
diode formed by two Ising-coupled qubits.
Each qubit is connected to an independent thermal reservoir and may be subject
to longitudinal periodic modulation.
Starting from a microscopic system--bath model, we derive a Floquet--LGKS
master equation that resolves the drive-assisted transition channels
by bath, Floquet sideband, and conditional Ising sector.
We use the resonant undriven device with left--right symmetric bath couplings
as the reciprocal benchmark under temperature reversal.
The known statically detuned configuration provides a rectifying reference in
which stronger asymmetry is accompanied by a reduced transmitted heat
current~\cite{PhysRevE.95.022128}. Single-side driving creates a Floquet-dressed contact opposite an undriven
thermal contact. This contact asymmetry produces directional current suppression.
For this configuration, we derive a compact factorized expression for the steady-state current and an exact blocking condition.
The condition can be satisfied for one bias direction while the reversed-bias current remains finite.
For weak sinusoidal driving with a flat bare spectrum, we further show that rectification emerges quadratically in the modulation amplitude.
For dual-side driving, we decompose the right-bath current into an
interaction-mediated cyclic contribution and a right-contact Floquet pumping
contribution. Complete blocking then requires cancellation of both terms. These results establish contact-selective Floquet dressing as the operating
principle of the diode and provide analytical criteria for directional heat-current blocking.

The remainder of the paper is organized as follows. In Sec.~\ref{sec:model}, we introduce the driven two-qubit model and derive the Floquet--LGKS master equation. Section~\ref{sec:results} presents the steady-state heat-transport analysis for the undriven, single-side-driven, and dual-side-driven regimes. We conclude in Sec.~\ref{sec:conclusions}. The appendices provide the microscopic derivation of the reduced dynamics and the analytical details underlying the steady-state current formulas.


\section{Model and Floquet Master Equation}\label{sec:model}

\begin{figure}[t]
\centering
\includegraphics[width=\columnwidth]{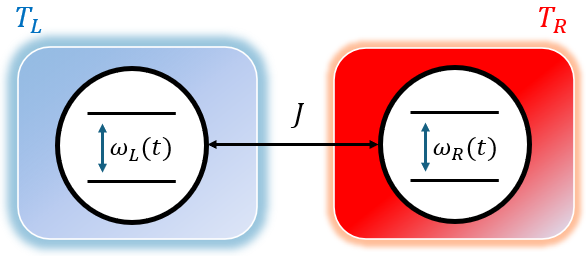}
\caption{Schematic of the Floquet quantum thermal diode. Two qubits with longitudinally modulated transition frequencies \(\omega_L(t)\) and \(\omega_R(t)\) are coupled through an Ising interaction \(J\) and connected to independent thermal reservoirs at temperatures \(T_L\) and \(T_R\). The same setup contains the undriven, single-side-driven, and dual-side-driven regimes analyzed in Sec.~\ref{sec:results}.}
\label{fig:schematic_setup}
\end{figure}

We consider a quantum thermal transport setup composed of two interacting qubits, labeled $L$ and $R$, each coupled weakly to an independent thermal reservoir. The qubits interact through an Ising coupling, while their transition frequencies may be modulated periodically in time by external longitudinal driving. A schematic representation of the device is shown in Fig.~\ref{fig:schematic_setup}. The total Hamiltonian is
\begin{equation}
H_{\mathrm{tot}}(t)
=
H_S(t)+H_B+\lambda_L H_{I,L}+\lambda_R H_{I,R},
\label{eq:Htot}
\end{equation}
where $H_S(t)$ denotes the driven two-qubit Hamiltonian, $H_B$ the bath Hamiltonian, and $H_{I,L}$ and $H_{I,R}$ the couplings to the left and right reservoirs. The dimensionless parameters $\lambda_L$ and $\lambda_R$ specify the weak system--bath coupling strengths.

The system Hamiltonian is given by
\begin{equation}
H_S(t)
=
\frac{\hbar}{2}\omega_L(t)\sigma_z^{(L)}
+
\frac{\hbar}{2}\omega_R(t)\sigma_z^{(R)}
+
J\,\sigma_z^{(L)}\sigma_z^{(R)},
\label{eq:HS_general}
\end{equation}
where $J$ is the Ising coupling strength. The time dependence enters through the longitudinally modulated qubit frequencies $\omega_L(t)$ and $\omega_R(t)$, which are assumed to be periodic with a common period $\tau$,
\begin{equation*}
\omega_L(t+\tau)=\omega_L(t),
\qquad
\omega_R(t+\tau)=\omega_R(t),
\end{equation*}
with driving frequency $\Omega=2\pi/\tau$. For later use, we decompose each qubit frequency into its cycle average and a purely periodic part,
\begin{equation}
\omega_a(t)=\omega_{a,0}+\delta\omega_a(t),
\qquad
a\in\{L,R\},
\label{eq:freq_decomposition}
\end{equation}
with
\begin{equation}
\frac{1}{\tau}\int_0^\tau \delta\omega_a(t)\,dt=0.
\label{eq:zero_mean_drive}
\end{equation}
Here $\omega_{a,0}$ denotes the cycle-averaged transition frequency, while $\delta\omega_a(t)$ contains the zero-mean periodic modulation. It is also convenient to introduce the Ising frequency scale
\begin{equation}
\Omega_J=\frac{2J}{\hbar},
\label{eq:OmegaJ}
\end{equation}
which will be used below to express the conditional transition frequencies generated by the interaction term.
The environment consists of two independent bosonic reservoirs,
\begin{equation}
H_B=H_B^{(L)}+H_B^{(R)},
\qquad
H_B^{(a)}=\sum_k \hbar\omega_{ak}\,b_{ak}^\dagger b_{ak},
\label{eq:HB}
\end{equation}
and the coupling of qubit $a$ to its reservoir is taken in factorized form,
\begin{equation}
H_{I,a}=S_a\otimes B_a,
\label{eq:HIa}
\end{equation}
We choose transverse system operators
\begin{equation}
S_L=\sigma_x^{(L)},
\qquad
S_R=\sigma_x^{(R)},
\label{eq:Sa}
\end{equation}
and bath operators
\begin{equation}
B_a=\sum_k g_{ak}\bigl(b_{ak}+b_{ak}^\dagger\bigr).
\label{eq:Ba}
\end{equation}
Thus, the external control acts longitudinally through $\sigma_z$, whereas dissipation acts transversely through $\sigma_x$ and induces qubit flips. Each reservoir is assumed to be in a thermal Gibbs state,
\begin{equation}
\rho_{B_a}
=
\frac{e^{-\beta_a H_B^{(a)}}}
{\mathrm{Tr}\!\left[e^{-\beta_a H_B^{(a)}}\right]},
\qquad
\beta_a=(k_B T_a)^{-1},
\label{eq:bath_states}
\end{equation}
so that the total bath state is $\rho_B=\rho_{B_L}\otimes\rho_{B_R}$. The two reservoirs are statistically independent and may in general be held at different temperatures $T_L$ and $T_R$.

Because the system Hamiltonian in Eq.~\eqref{eq:HS_general} is diagonal in the computational basis $ \left\{ \ket{e_Le_R}, \ket{e_Lg_R}, \ket{g_Le_R}, \ket{g_Lg_R} \right\}$, its instantaneous eigenenergies are
\begin{align}
E_{ee}(t)
&=
+\frac{\hbar}{2}\omega_L(t)
+\frac{\hbar}{2}\omega_R(t)
+J,
\nonumber\\
E_{eg}(t)
&=
+\frac{\hbar}{2}\omega_L(t)
-\frac{\hbar}{2}\omega_R(t)
-J,
\nonumber\\
E_{ge}(t)
&=
-\frac{\hbar}{2}\omega_L(t)
+\frac{\hbar}{2}\omega_R(t)
-J,
\nonumber\\
E_{gg}(t)
&=
-\frac{\hbar}{2}\omega_L(t)
-\frac{\hbar}{2}\omega_R(t)
+J.
\label{eq:instantaneous_energies}
\end{align}
These expressions show that the energy cost of flipping one qubit depends on the state of the other, so that the Ising interaction resolves each bare qubit transition into two conditional sectors. Accordingly, when the right qubit is fixed in the sector $s_R=\pm1$, the transition frequency of the left qubit is
\begin{equation}
\nu_L^{(s_R)}(t)=\omega_L(t)+\Omega_J s_R,
\label{eq:nuL_conditional}
\end{equation}
whereas, when the left qubit is fixed in the sector $s_L=\pm1$, the transition frequency of the right qubit is
\begin{equation}
\nu_R^{(s_L)}(t)=\omega_R(t)+\Omega_J s_L.
\label{eq:nuR_conditional}
\end{equation}
In the static limit, these reduce to the familiar sector-resolved transition frequencies of the Ising-coupled two-qubit system \cite{PhysRevE.95.022128}. Under periodic driving, they become time-periodic conditional frequencies and provide the basic frequencies from which the Floquet sideband structure of the reduced dynamics is built. Throughout the remainder of the work, we use the ladder and projector operators
\begin{equation}
\sigma_\pm^{(a)}
=
\frac{1}{2}\bigl(\sigma_x^{(a)}\pm i\sigma_y^{(a)}\bigr),
\qquad
\Pi_\pm^{(a)}
=
\frac{1}{2}\bigl(\mathbbm{1}\pm\sigma_z^{(a)}\bigr),
\label{eq:operator_conventions}
\end{equation}
with $a\in\{L,R\}$. Here $\Pi_+^{(a)}$ and $\Pi_-^{(a)}$ project onto the excited and ground sectors of qubit $a$, respectively. These operators provide a convenient basis for constructing the interaction-picture transition operators and for resolving the Floquet master equation into sector- and sideband-resolved channels.

\subsection{Floquet decomposition and reduced master equation}

We now derive the reduced driven dynamics in the form needed for the transport analysis. Starting from the microscopic model defined in Eqs.~\eqref{eq:Htot}--\eqref{eq:Ba}, we move to the interaction picture with respect to the driven system Hamiltonian \(H_S(t)\) in Eq.~\eqref{eq:HS_general}. Because \(H_S(t)\) remains diagonal at all times, the interaction-picture system operators can be decomposed into channels labeled by both a Floquet sideband index and a conditional-sector index. The Floquet structure originates from the periodic longitudinal modulation, while the sector structure reflects the Ising-induced splitting of the qubit transition frequencies. As shown explicitly in Appendix~\ref{app:Floquet_LGKS_derivation}, this decomposition provides the natural basis for constructing the reduced dissipative dynamics. Substituting the resulting interaction-picture operators into the two-bath Redfield equation~\eqref{eq:app_Redfield}, and then applying the Born, Markov, and Floquet--secular approximations, yields a time-local Floquet--LGKS master equation in the interaction picture \cite{Floquet1883, breuer2002, Szczygielski2013, GelbwaserKlimovsky2015, Binder2018},
\begin{equation}
\begin{aligned}
\frac{d}{dt}\rho_{S,I}(t)
&=
-\frac{i}{\hbar}[H_{\mathrm{LS}},\rho_{S,I}(t)]
\\
&\quad
+\sum_{q,\sigma}
\Gamma_{\downarrow}^{(L)}(q,\sigma)\,
\mathcal D\!\bigl[L_{q,\sigma,\downarrow}^{(L)}\bigr]\rho_{S,I}(t)
\\
&\quad+\sum_{q,\sigma}
\Gamma_{\uparrow}^{(L)}(q,\sigma)\,
\mathcal D\!\bigl[L_{q,\sigma,\uparrow}^{(L)}\bigr]\rho_{S,I}(t)
\\
&\quad
+\sum_{p,\sigma}
\Gamma_{\downarrow}^{(R)}(p,\sigma)\,
\mathcal D\!\bigl[L_{p,\sigma,\downarrow}^{(R)}\bigr]\rho_{S,I}(t)
\\
&\quad+\sum_{p,\sigma}
\Gamma_{\uparrow}^{(R)}(p,\sigma)\,
\mathcal D\!\bigl[L_{p,\sigma,\uparrow}^{(R)}\bigr]\rho_{S,I}(t).
\end{aligned}
\label{eq:main_Floquet_LGKS_IP}
\end{equation}
Here \(\rho_{S,I}(t)\) denotes the reduced density operator in the interaction picture, and
\begin{equation*}
\mathcal D[L]\rho
=
L\rho L^\dagger
-
\frac{1}{2}\{L^\dagger L,\rho\}
\end{equation*}
is the standard Lindblad dissipator. The coherent correction \(H_{\mathrm{LS}}\) is the Lamb-shift Hamiltonian, whose explicit form is given in Eq.~\eqref{eq:app_HLS}.
The indices \(q\) and \(p\) label the Floquet sidebands associated with the left and right periodic modulations, respectively, while \(\sigma=\pm\) resolves the two conditional sectors generated by the Ising splitting. The corresponding jump operators are listed in Eqs.~\eqref{eq:app_JumpsL} and \eqref{eq:app_JumpsR}, and the associated transition rates in Eqs.~\eqref{eq:app_RatesL} and \eqref{eq:app_RatesR}.

Equation~\eqref{eq:main_Floquet_LGKS_IP} is the central dynamical result of the microscopic construction. It shows that periodic driving resolves each bath-induced transition into a ladder of Floquet sidebands, while the Ising interaction further splits each sideband into two conditional sectors. The reduced dynamics therefore acquire a channel structure that is resolved simultaneously by bath, sideband, and sector, which is the key feature underlying the transport asymmetries analyzed below. The single-side-driven and undriven limits follow directly by suppressing one or both modulation functions in Eq.~\eqref{eq:freq_decomposition}; in those limits, the corresponding Floquet ladders collapse to their central components.

\section{Results}
\label{sec:results}

Having derived the Floquet--LGKS master equation, we now analyze the steady-state heat-transport response of the device. 
This section is organized around three cases. We first use the symmetric undriven configuration as a reference point,
where the resonant static device is known to be reciprocal and therefore shows no thermal rectification \cite{PhysRevE.95.022128}. 
We then turn to the single-side-driven configuration, which contains the main Floquet-induced rectification mechanism of this work.
Finally, we examine the dual-side-driven case to determine how adding a second modulation modifies the transport asymmetry of the device.

\subsection{Heat currents and figures of merit}
\label{subsec:heat_currents_figures_merit}

To define heat currents in a form consistent with the Floquet decomposition, we use the cycle-averaged system Hamiltonian as the 
energy observable for steady-state transport,
\begin{align}
H_F
&=
\frac{1}{\tau}\int_0^\tau H_S(t)\,dt \nonumber\\
&=
\frac{\hbar}{2}\omega_{L,0}\sigma_z^{(L)}
+
\frac{\hbar}{2}\omega_{R,0}\sigma_z^{(R)}
+
J\,\sigma_z^{(L)}\sigma_z^{(R)} ,
\label{eq:results_floquet_hamiltonian}
\end{align}
where $\omega_{L,0}$ and $\omega_{R,0}$ are the cycle-averaged qubit frequencies defined in Eq.~\eqref{eq:freq_decomposition}. 
If the dissipative generator associated with reservoir $a$ is decomposed into Floquet channels labeled by the quasienergy gap $\omega$
and sideband index $q$, the steady-state heat current from that reservoir is \cite{GelbwaserKlimovsky2015, PhysRevE.87.012140, Gupt2022}
\begin{equation}
\mathcal{J}_a^{\mathrm{ss}}
=
\sum_{\omega,q}
\frac{\omega+q\Omega}{\omega}\,
\operatorname{Tr}\!\left[
\mathcal{L}_{\omega,q}^{(a)}
\!\left[\rho_{\mathrm{ss}}\right] H_F
\right].
\label{eq:results_heat_current_definition}
\end{equation}
The prefactor $(\omega+q\Omega)/\omega$ accounts for the fact that a Floquet transition with quasienergy gap $\omega$ exchanges
the physical energy $\hbar(\omega+q\Omega)$ with the reservoir through sideband $q$. In this way, the heat current counts the energy
carried by each reservoir-induced transition, including the energy supplied or absorbed through the periodic drive.

For an undriven contact, only the central sideband contributes. Setting $q=0$ in Eq.~\eqref{eq:results_heat_current_definition}
gives \cite{PhysRevE.66.036102,GelbwaserKlimovsky2015,Gupt2022}
\begin{equation}
\mathcal{J}_a^{\mathrm{ss}}
=
\sum_{\omega}
\operatorname{Tr}\!\left[
\mathcal{L}_{\omega,0}^{(a)}
\!\left[\rho_{\mathrm{ss}}\right] H_F
\right].
\label{eq:results_heat_current_undriven}
\end{equation}
We use the sign convention that $\mathcal{J}_a^{\mathrm{ss}}>0$ means reservoir $a$ injects energy into the system,
while $\mathcal{J}_a^{\mathrm{ss}}<0$ means that energy flows from the system into that reservoir. The forward and reverse thermal biases
are defined as
\begin{equation}
(T_L,T_R)=(T_h,T_c),
\qquad
(T_L,T_R)=(T_c,T_h),
\label{eq:results_bias_convention}
\end{equation}
respectively, with $T_h>T_c$. Throughout the work, the steady-state current exchanged with the right reservoir is used to quantify diode performance. For the forward and reverse thermal biases defined above, this gives
\begin{equation}
\mathcal{J}_F
\equiv
\mathcal{J}_R^{\mathrm{ss}}(T_h,T_c),
\qquad
\mathcal{J}_B
\equiv
\mathcal{J}_R^{\mathrm{ss}}(T_c,T_h).
\label{eq:results_forward_reverse_currents}
\end{equation}

We characterize the resulting directional asymmetry using two complementary measures. The rectification coefficient is defined as \cite{PhysRevE.99.042121}
\begin{equation}
\mathcal{R}
=
\frac{|\mathcal{J}_F+\mathcal{J}_B|}
{\max(|\mathcal{J}_F|,|\mathcal{J}_B|)} .
\label{eq:results_rectification_coefficient}
\end{equation}
This quantity vanishes when reversing the thermal bias changes only the sign of the current, not its magnitude. As a second measure, we use the rectification ratio \cite{Senior2020,RevModPhys.93.041001}
\begin{equation}
\eta
=
\frac{|\mathcal{J}_F|}{|\mathcal{J}_B|}.
\label{eq:results_rectification_ratio}
\end{equation}
For nonzero forward and reverse currents, the reciprocal limit is therefore characterized by
\begin{equation*}
\mathcal{R}=0,
\qquad
|\mathcal{J}_F|=|\mathcal{J}_B|,
\qquad
\eta=1 .
\end{equation*}
Therefore, departures from this limit indicate directional asymmetry under reversal of the temperature bias.

Throughout the transport analysis, we assume that the bare reservoir spectrum
is flat over the range of Bohr and Floquet-sideband frequencies relevant to the
system, as specified in Eq.~\eqref{eq:undriven_flat_spectrum}.
This wide-band approximation is commonly used in weak-coupling Markovian
treatments when the reservoir spectrum varies slowly across the transition
frequencies sampled by the system \cite{breuer2002,McCauley2020}. The assumption concerns only the bare system--reservoir coupling spectrum.
The complete excitation and relaxation rates remain frequency dependent through
the thermal occupation factors evaluated at the corresponding transition
frequencies. Using a flat bare spectrum therefore removes additional frequency filtering by
the reservoirs and allows us to isolate the transport asymmetry generated by
the Floquet sidebands and the Ising-resolved transition sectors.
Similar simplifying assumptions have been adopted in microscopic studies of
driven quantum thermal machines and heat transport
\cite{PhysRevE.87.012140,GelbwaserKlimovsky2015,PhysRevE.99.042121,PhysRevE.104.054137}.
The approximation nonetheless has a limited range of validity because realistic
reservoirs may exhibit finite cutoffs, resonant peaks, band gaps, or other
frequency-dependent structures. Such spectral features can reweight the sideband-resolved transition rates and
thereby modify both the magnitude and the rectification of the heat current.
Reservoir spectral engineering may therefore provide an additional means of
controlling heat transport
\cite{PhysRevResearch.2.033285}. A systematic investigation of structured reservoir spectra goes beyond the
scope of the current work.

\subsection{Undriven reference case}
\label{subsec:undriven_reference_case}

Before turning to the periodically driven configurations, we first recall the undriven static limit as a benchmark. This case is not presented as a new result: the transport properties of the undriven Ising-coupled two-qubit model have been analyzed previously \cite{PhysRevE.95.022128}. Its purpose here is to establish the reciprocal reference point against which the Floquet-driven regimes will be compared.

Setting $\delta\omega_L(t)=\delta\omega_R(t)=0$ in Eq.~\eqref{eq:freq_decomposition} removes all Floquet sidebands and reduces the dynamics to the static sector-resolved LGKS equation given in Appendix~\ref{app:undriven_normalization_factor_appendix}. Solving the corresponding steady-state rate equations and substituting the result into the heat-current definition in Eq.~\eqref{eq:results_heat_current_undriven} gives
\begin{equation}
\mathcal{J}_R^{\mathrm{ss}}
=
\frac{2\hbar\Omega_J}{\mathcal Z}
\prod_{a\in\{L,R\}}
\prod_{\sigma=\pm}
\Gamma_{\downarrow}^{(a,\sigma)}
\bigl[
r_L^{(-)}r_R^{(+)}
-
r_L^{(+)}r_R^{(-)}
\bigr],
\label{eq:undriven_right_current_final}
\end{equation}
where the relaxation rates $\Gamma_{\downarrow}^{(a,\sigma)}$ and the detailed-balance ratios $r_a^{(\sigma)}$ are defined in Eqs.~\eqref{eq:undriven_rate_definitions} and \eqref{eq:undriven_DB_rates}, respectively. The normalization factor $\mathcal Z>0$ is given in Eq.~\eqref{eq:app_undriven_normalization_factor}. Thus, apart from the fixed prefactor, the sign of the current is controlled entirely by the bracketed detailed-balance combination.

The reciprocal benchmark becomes transparent in the resonant left--right symmetric limit, $\omega_L=\omega_R\equiv\omega_0, \, \gamma_L=\gamma_R$. In this case the bracket in Eq.~\eqref{eq:undriven_right_current_final} reduces to
\begin{equation}
r_L^{(-)}r_R^{(+)}-r_L^{(+)}r_R^{(-)}
=
2e^{-(\beta_L+\beta_R)\hbar\omega_0}
\sinh\!\big[(\beta_L-\beta_R)\hbar\Omega_J\big].
\label{eq:undriven_current_bracket_symmetric}
\end{equation}
This expression is odd under temperature reversal, $T_L\leftrightarrow T_R$. Since the normalization factor in Eq. \eqref{eq:undriven_right_current_final} remains unchanged under the same exchange, the right-bath current satisfies
\begin{equation}
\mathcal{J}_R^{\mathrm{ss}}(T_L,T_R)
=
-
\mathcal{J}_R^{\mathrm{ss}}(T_R,T_L).
\label{eq:undriven_current_antisymmetry}
\end{equation}
It follows that
\begin{equation*}
|\mathcal{J}_F|=|\mathcal{J}_B|,
\qquad
\mathcal R=0,
\qquad
\eta=1.
\end{equation*}

The symmetric resonant undriven device is therefore reciprocal: reversing the temperature bias changes only the direction of the heat current, not its magnitude. Static rectification can be obtained by moving away from the resonant symmetric limit, for example by taking $\omega_L\neq\omega_R$. However, this involves a tradeoff between stronger rectification and reduced transmitted current \cite{PhysRevE.95.022128}. This limitation provides the motivation for the periodically driven configurations considered next.

\subsection{Single-side driving}
\label{subsec:single_side_driving}

We now consider the central driven configuration of this work: only the left qubit is longitudinally modulated, while the right qubit remains static. In the notation of Eq.~\eqref{eq:freq_decomposition}, this specialization is
\begin{equation}
\delta\omega_L(t)\neq 0,
\qquad
\delta\omega_R(t)=0 .
\label{eq:single_side_drive_specialization}
\end{equation}
This is the minimal driven geometry that can break the reciprocal structure of the undriven benchmark in a controlled way. In this case, the left contact acquires Floquet sidebands, whereas the right contact retains only the two static sector-resolved thermal channels as shown in Eq. \eqref{eq:app_single_side_generator}. For each right-qubit sector $\sigma=\pm$, we characterize the driven left contact by the effective Floquet ratio
\begin{equation}
r_\sigma^{(L)}(T_L)
=
\frac{
\sum_q \Gamma_{\uparrow}^{(L)}(q,\sigma;T_L)
}{
\sum_q \Gamma_{\downarrow}^{(L)}(q,\sigma;T_L)
}.
\label{eq:single_side_rsigma_def}
\end{equation}
The numerator and denominator are, respectively, the total sideband-assisted excitation and relaxation rates of the left qubit in sector $\sigma$.

With these effective ratios, the steady-state right-bath heat current can be written in the factorized form
\begin{equation}
\mathcal{J}_R^{\mathrm{ss}}(T_L,T_R)
=
2\hbar\Omega_J\,
\mathcal K(T_L,T_R)\,
\Xi(T_L,T_R).
\label{eq:single_side_factorized_current}
\end{equation}
A detailed derivation of Eq.~\eqref{eq:single_side_factorized_current} is presented in Appendix~\ref{appsubsec:single_side_current_derivation}. The prefactor $\mathcal K(T_L,T_R)>0$, given explicitly in Eq.~\eqref{eq:app_single_side_K}, contains only positive transition-rate and normalization factors. Therefore, the sign and zeros of the current are determined entirely by
\begin{equation}
\Xi(T_L,T_R)
=
r_-^{(L)}(T_L)e^{-\beta_R\hbar \nu_R^{(+)}}
-
r_+^{(L)}(T_L)e^{-\beta_R\hbar \nu_R^{(-)}},
\label{eq:single_side_Xi_def}
\end{equation}
where $\nu_R^{(\pm)}$ are the right bath transition frequencies and given in Eq. \eqref{eq:app_single_side_right_freqs}.
Equation~\eqref{eq:single_side_Xi_def} identifies the mechanism of single-side-driven rectification. The undriven right contact contributes ordinary sector-resolved Boltzmann factors, $e^{-\beta_R\hbar \nu_R^{(\pm)}}$, while the driven left contact contributes the Floquet-renormalized ratios $r_\sigma^{(L)}$. Because these ratios combine all sideband-assisted excitation and relaxation pathways, they need not transform under temperature reversal like equilibrium detailed-balance factors. The resulting mismatch between the Floquet-dressed left contact and the thermal right contact produces directional transport asymmetry.

\begin{figure*}[t]
\centering
\subfigure[]{
\includegraphics[width=0.32\textwidth]{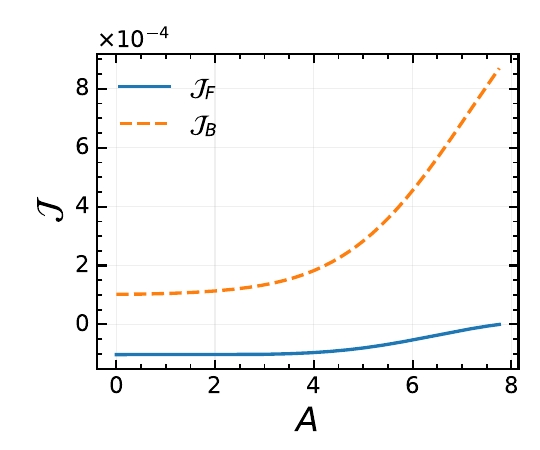}
\label{fig:fig_2a}
}
\hfill
\subfigure[]{
\includegraphics[width=0.32\textwidth]{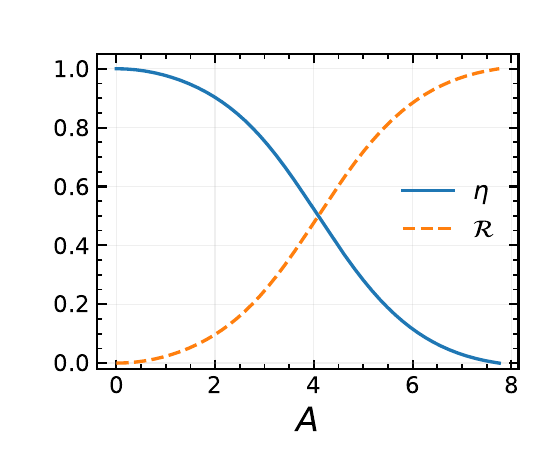}
\label{fig:fig_2b}
}
\hfill
\subfigure[]{
\includegraphics[width=0.32\textwidth]{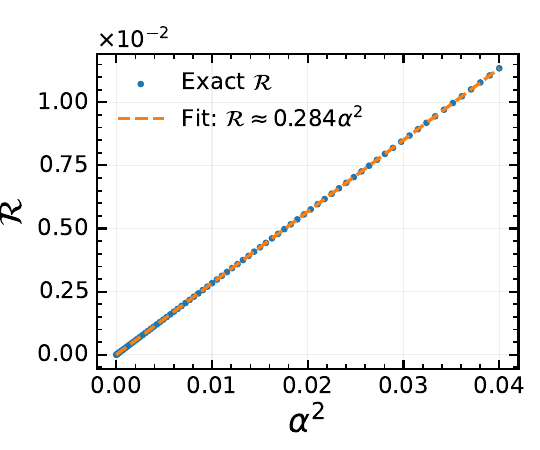}
\label{fig:fig_2c}
}
\caption{Numerical verification of single-side-driven rectification using the sinusoidal left-qubit modulation in Eq.~\eqref{eq:single_side_sinusoidal_drive} and the flat bare reservoir spectrum in Eq.~\eqref{eq:undriven_flat_spectrum}.
(a) Steady-state right-bath currents $\mathcal{J}_F$ and $\mathcal{J}_B$, defined in Eq.~\eqref{eq:results_forward_reverse_currents}, versus the drive amplitude $A$ for the forward and reverse thermal biases in Eq.~\eqref{eq:results_bias_convention}. Increasing $A$ selectively suppresses the forward current, which vanishes at the first blocking point, $A_{\mathrm{block}}\simeq 7.754$.
(b) Rectification ratio $\eta$ and rectification coefficient $\mathcal{R}$, defined in Eqs.~\eqref{eq:results_rectification_ratio} and \eqref{eq:results_rectification_coefficient}, respectively. At the forward-current blocking point, $\eta\rightarrow0$ and $\mathcal{R}\rightarrow1$, demonstrating nearly ideal rectification while the reverse transport branch remains conducting.
(c) Weak-drive verification of the quadratic scaling predicted by Eq.~\eqref{eq:single_side_weak_drive_scaling}, shown as a function of $\alpha^2$, with $\alpha$ defined in Eq.~\eqref{eq:single_side_alpha_def}. 
The parameters are $\omega_{L,0}=\omega_R=12$, $J=0.55$, $\Omega=3.5$, $T_h=8$, $T_c=4$, and $\kappa_L=\kappa_R=0.04$, with $\hbar=k_B=1$. The Floquet sums are evaluated using symmetric, convergence-controlled sideband truncation.}
\label{fig:fig2}
\end{figure*}

The same factorized structure of Eq. \eqref{eq:single_side_factorized_current} also gives a direct heat current blocking criterion. Away from disconnected limits, $\mathcal K(T_L,T_R)>0$, so the current vanishes exactly when $\Xi(T_L,T_R)=0$. Using the sector splitting $\nu_R^{(+)}-\nu_R^{(-)}=2\Omega_J$ and Eq. \eqref{eq:single_side_Xi_def}, this blocking condition can be expressed through
\begin{equation}
\Theta(T_L,T_R)
=
\ln\!\frac{r_-^{(L)}(T_L)}{r_+^{(L)}(T_L)}
-
2\beta_R\hbar\Omega_J .
\label{eq:single_side_control_exponent}
\end{equation}
Thus,
\begin{equation}
\mathcal{J}_R^{\mathrm{ss}}(T_L,T_R)=0
\qquad
\Longleftrightarrow
\qquad
\Theta(T_L,T_R)=0 .
\label{eq:single_side_zero_current_condition}
\end{equation}
The condition $\Theta=0$ states that current blocking occurs when the Floquet-induced sector asymmetry of the driven left contact exactly compensates the thermal sector splitting imposed by the static right contact. Using the forward and reverse bias conventions introduced in Eq.~\eqref{eq:results_bias_convention}, the corresponding blocking parameters are
\begin{align}
\Theta_{F,B}
&=
\ln\!\frac{r_-^{(L)}(T_{h, c})}{r_+^{(L)}(T_{h, c})}
-
2\beta_{c, h}\hbar\Omega_J.
\label{eq:single_side_forward_blocking}
\end{align}
Forward blocking occurs when $\Theta_F=0$ while $\Theta_B\neq0$, whereas reverse blocking occurs when $\Theta_B=0$ while $\Theta_F\neq0$. These conditions provide a direct analytical design rule for strong rectification: one bias direction can be tuned close to its blocking point while the opposite direction remains conducting.

To connect the general blocking condition with the model used in the numerical analysis, we now specialize to sinusoidal modulation of the left qubit while keeping the right qubit static. The driven frequency is taken as
\begin{equation}
\omega_L(t)
=
\omega_{L,0}
+
A\cos(\Omega t),
\label{eq:single_side_sinusoidal_drive}
\end{equation}
where $A$ is the modulation amplitude and $\Omega$ is the drive frequency. For this sinusoidal modulation, the general sideband-weight definition in Eq.~\eqref{eq:app_Pa} reduces to
\begin{equation}
P(q)
=
\mathrm J_q^2\!\left(\frac{A}{\Omega}\right).
\label{eq:single_side_sideband_weights}
\end{equation}
Here $\mathrm J_q$ is the Bessel function of the first kind. The corresponding left-contact Bohr--Floquet frequencies are
\begin{equation}
\nu_{L,q}^{(\sigma)}
=
\omega_{L,0}+q\Omega+\sigma\Omega_J .
\label{eq:single_side_nuLq}
\end{equation}
The term $q\Omega$ denotes the energy shift associated with the $q$th
Floquet sideband, while $\sigma\Omega_J$ is the conditional shift
generated by the Ising interaction. Because the complete Floquet ladder
may contain both positive- and negative-frequency channels, the
flat-spectrum rates must be evaluated using the corresponding branches
of Eq.~\eqref{eq:undriven_flat_spectrum}. The resulting effective
occupations, frequency-branch weights, and driven-contact ratios are
defined explicitly in Eqs.~\eqref{eq:app_single_side_Nsigma}--%
\eqref{eq:app_single_side_flat_rsigma}. The general rate-based
definition in Eq.~\eqref{eq:single_side_rsigma_def} remains valid
independently of the signs of the individual sideband frequencies.
For the weak-drive analysis considered below, only the carrier and first
sidebands contribute through second order in the modulation amplitude.
If $\omega_{L,0}>\Omega+|\Omega_J|$, all channels retained at this order
lie on the positive-frequency branch. The driven-contact ratio then
reduces to
\begin{equation}
r_\sigma^{(L)}(T_L)
=
\frac{N_\sigma(T_L)}
{1+N_\sigma(T_L)}
+
\mathcal O\!\left[
\left(\frac{A}{\Omega}\right)^4
\right].
\label{eq:single_side_rsigma_flat}
\end{equation}
Substitution of the exact signed-frequency ratios into
Eq.~\eqref{eq:single_side_factorized_current} yields the flat-spectrum
current derived in Eq.~\eqref{eq:app_single_side_flat_factorized}. Its
factorized structure remains unchanged: the kinetic prefactor is
positive, while the competition bracket determines the current
direction and blocking points. The sinusoidal drive modifies these
quantities by redistributing the left-contact Floquet weights between
the two conditional sectors.

For the sinusoidal drive in Eq.~\eqref{eq:single_side_sinusoidal_drive}, the rectification coefficient defined in Eq.~\eqref{eq:results_rectification_coefficient} can be evaluated from the full factorized current. At arbitrary drive strength, however, the resulting expression contains many sideband contributions and is too cumbersome to provide a transparent design rule. We therefore focus on the weak-drive regime, where the onset of rectification can be obtained analytically near the reciprocal undriven benchmark. The relevant expansion parameter is
\begin{equation}
\alpha
=
\frac{A}{\Omega},
\qquad
\alpha\ll1 .
\label{eq:single_side_alpha_def}
\end{equation}
For sinusoidal modulation, the sideband weights in Eq.~\eqref{eq:single_side_sideband_weights} have the leading expansion
\begin{equation}
P(0)
=
1-\frac{\alpha^2}{2}
+
\mathcal{O}(\alpha^4),
\qquad
P(\pm1)
=
\frac{\alpha^2}{4}
+
\mathcal{O}(\alpha^4),
\label{eq:single_side_weak_sidebands}
\end{equation}
while higher sidebands enter only at order $\mathcal{O}(\alpha^4)$. Thus, the first drive-induced correction to the effective occupations, and hence to the ratios $r_\sigma^{(L)}$, appears at order $\alpha^2$. We now consider the symmetric static reference point, $\omega_{L,0}=\omega_R\equiv\omega_0, \,\gamma_L=\gamma_R$, for which the undriven current is reciprocal and gives no rectification. Expanding the sinusoidal flat-spectrum current around this reference point gives the weak-drive scaling
\begin{equation}
\mathcal R
=
\alpha^2
\mathcal F(T_h,T_c,\omega_0,J,\Omega)
+
\mathcal{O}(\alpha^4).
\label{eq:single_side_weak_drive_scaling}
\end{equation}
The derivation is given in Appendix~\ref{appsubsec:single_side_weak_drive}, where the coefficient $\mathcal F$ is defined explicitly in Eq.~\eqref{eq:app_single_side_Fcoef}. Equation~\eqref{eq:single_side_weak_drive_scaling} shows that rectification turns on as the leading Floquet correction to the reciprocal undriven device. The absence of a linear term follows from the even dependence of the Bessel sideband weights on the modulation amplitude. The analytical results above provide two complementary benchmarks for the numerical analysis. The exact factorized current gives the blocking criteria in Eqs.~\eqref{eq:single_side_zero_current_condition}--\eqref{eq:single_side_forward_blocking}, while the sinusoidal flat-spectrum specialization gives the weak-drive scaling law in Eq.~\eqref{eq:single_side_weak_drive_scaling}. We now compare these predictions with numerical steady-state transport data.

Figure~\ref{fig:fig2} compares the numerical steady-state results with two analytical predictions of the single-side-driven model. Panels~\ref{fig:fig_2a} and \ref{fig:fig_2b} test the current-blocking conditions in Eqs.~\eqref{eq:single_side_zero_current_condition}--\eqref{eq:single_side_forward_blocking}, whereas panel~\ref{fig:fig_2c} tests the quadratic weak-drive scaling in Eq.~\eqref{eq:single_side_weak_drive_scaling}. The three panels thus connect the analytical blocking mechanism with the current branches, rectification measures, and perturbative weak-drive behavior.
At $A=0$, Fig.~\ref{fig:fig_2a} recovers the symmetric undriven reference case discussed in Sec.~\ref{subsec:undriven_reference_case}. The forward and reverse currents have equal magnitudes and opposite signs, consistent with the reciprocity relation in Eq.~\eqref{eq:undriven_current_antisymmetry}. Driving the left qubit redistributes the sideband-resolved transition weights and thereby modifies the effective ratios $r_\sigma^{(L)}$ in Eq.~\eqref{eq:single_side_Xi_def}. The undriven right contact retains its thermal detailed-balance factors. This contact asymmetry suppresses the forward current relative to the reverse current. At the first forward-blocking point, $A_{\mathrm{block}}\simeq 7.754$, the numerical current satisfies $\mathcal{J}_F\simeq -4.1\times10^{-17}$, while the reverse branch remains conducting with $\mathcal{J}_B\simeq 8.69\times10^{-4}$. The corresponding blocking parameters are $\Theta_F\simeq 3.3\times10^{-16}$ and $\Theta_B\simeq 0.732$, in quantitative agreement with the criterion in Eq.~\eqref{eq:single_side_forward_blocking}.

Figure~\ref{fig:fig_2b} summarizes the resulting directional asymmetry. The undriven values $\mathcal{R}=0$ and $\eta=1$ define the reciprocal reference point. As the drive approaches the forward-blocking point, the two measures evolve toward $\mathcal{R}=1$ and $\eta=0$, respectively. Because the reverse current remains of order $10^{-3}$, the large rectification coefficient does not result from the simultaneous suppression of both transport branches. Moreover, the currents remain oppositely directed throughout the open interval $0\leq A<A_{\mathrm{block}}$, for which $\mathcal{J}_F\mathcal{J}_B<0$; equality occurs only at the blocking endpoint. The plotted range therefore describes a diode-like regime rather than a regime in which both thermal biases produce currents in the same direction through drive-induced pumping.

Figure~\ref{fig:fig_2c} provides a quantitative test of the weak-drive expansion. Over $0\leq\alpha\leq 0.2$, the numerical rectification coefficient is linear in $\alpha^2$, as predicted by Eq.~\eqref{eq:single_side_weak_drive_scaling}. A fit through the origin gives a coefficient of $0.283647$, compared with the analytical value $\mathcal{F}=0.283568$ obtained from Eq.~\eqref{eq:app_single_side_Fcoef}. This agreement supports the perturbative result that rectification first appears at order $\alpha^2$. As indicated by the sideband expansion in Eq.~\eqref{eq:single_side_weak_sidebands}, the absence of a term linear in $\alpha$ follows from the even dependence of the sinusoidal sideband weights on the modulation amplitude.

Taken together, these results identify the operating principle of the single-side-driven diode. The Ising interaction separates transport into sector-dependent channels, and the drive redistributes the left-contact sideband weights so that one bias direction can be tuned toward a blocking point. Strong rectification is obtained when this suppression occurs without closing the opposite transport branch. This gives a simple design rule for Floquet-controlled heat rectification: engineer directional spectral mismatch at one contact while keeping the other contact thermal and conducting. We next examine how this picture changes when both qubits are modulated.

\subsection{Dual-side driving}
\label{subsec:dual_side_driving}

We now extend the single-side-driven configuration by modulating both qubits longitudinally. This modification changes the structure of the measured right-bath heat current because the right contact, where the current is evaluated, is itself Floquet dressed. As a result, the current contains two physically distinct contributions: an interaction-mediated cyclic contribution associated with the stationary population flow around the four-state loop, and a right-contact Floquet pumping contribution generated by sideband energy exchange at the measured contact.

We consider sinusoidal modulation of both transition frequencies with a common driving frequency,
\begin{equation}
\omega_a(t)
=
\omega_{a,0}
+
A_a\cos(\Omega t).
\label{eq:dual_side_drive_protocol}
\end{equation}
The single-side-driven configuration discussed in
Sec.~\ref{subsec:single_side_driving} is recovered by setting \(A_R=0\).
For \(A_R\neq0\), both contacts acquire Floquet sidebands. We define the
sideband-resolved upward and downward rates by
\begin{align}
u_{a,n}^{\sigma}
& =
\Gamma_{\uparrow}^{(a)}(n,\sigma)
=
\frac{\lambda_a^2}{\hbar^2}
P_a(n)
G_a\!\left(-\nu_{a,n}^{(\sigma)}\right),
\nonumber\\
d_{a,n}^{\sigma}
& =
\Gamma_{\downarrow}^{(a)}(n,\sigma)
=
\frac{\lambda_a^2}{\hbar^2}
P_a(n)
G_a\!\left(\nu_{a,n}^{(\sigma)}\right),
\nonumber\\
u_a^\sigma
&=
\sum_{n\in\mathbb Z}u_{a,n}^{\sigma},
\qquad
d_a^\sigma
=
\sum_{n\in\mathbb Z}d_{a,n}^{\sigma}.
\label{eq:dual_side_effective_rates}
\end{align}
The bath spectral function in these expressions is evaluated using the
appropriate signed-frequency branch of
Eq.~\eqref{eq:undriven_flat_spectrum}. Exact zero-frequency channels are
excluded. Channel by channel, the KMS relation gives
\begin{equation*}
\frac{u_{a,n}^{\sigma}}{d_{a,n}^{\sigma}}
=
e^{-\beta_a\hbar\nu_{a,n}^{(\sigma)}} .
\end{equation*}

Applying the Floquet heat-current definition in Eq.~\eqref{eq:results_heat_current_definition} to the Floquet--LGKS dynamics in Eq.~\eqref{eq:main_Floquet_LGKS_IP}, and using the stationary four-state solution derived in Appendix~\ref{appsubsec:dual_side_current}, gives the right-bath current
\begin{equation}
\mathcal{J}_R^{\rm ss}
=
\frac{\hbar}{Z_{\rm dd}}
\left[
2\Omega_J
\left(
\mathcal{C}_{\rm f}
-
\mathcal{C}_{\rm b}
\right)
+
\Omega\mathcal{B}_R
\right].
\label{eq:dual_side_factorized_current}
\end{equation}
The first term in Eq.~\eqref{eq:dual_side_factorized_current} is the interaction-mediated cyclic contribution. It is proportional to the imbalance \(\mathcal{C}_{\rm f}-\mathcal{C}_{\rm b}\) between the forward and backward products of rates around the four-state population loop. These products are defined explicitly in Eq.~\eqref{eq:app_dual_cycle_current_rate_form}, and the positive denominator \(Z_{\rm dd}\), defined in Eq.~\eqref{eq:app_dual_Zdd_definition}, normalizes the stationary solution. Thus, this contribution is controlled by the direction and magnitude of the net stationary probability flux around the population cycle. 

The second term in Eq.~\eqref{eq:dual_side_factorized_current} is the right-contact Floquet pumping contribution. The pumping numerator \(\mathcal{B}_R\) is defined in Eq.~\eqref{eq:app_dual_BR_definition} in terms of the sideband-weighted right-contact rates and the unnormalized stationary populations \(Q_X\) listed in Eq.~\eqref{eq:app_dual_Q_definitions}. Unlike the cyclic contribution, this term is present only when the measured right contact is driven. In particular, it vanishes for \(A_R=0\), because then only the \(q=0\) sideband contributes and all first sideband moments at the right contact are zero. Because \(Z_{\rm dd}>0\), Eq.~\eqref{eq:dual_side_factorized_current} shows that the full right-bath heat current vanishes when
\begin{equation}
2\Omega_J
\left(
\mathcal{C}_{\rm f}
-
\mathcal{C}_{\rm b}
\right)
+
\Omega\mathcal{B}_R
=
0 .
\label{eq:dual_side_blocking_condition}
\end{equation}
This condition is stronger than the cycle-balance condition \(\mathcal{C}_{\rm f}=\mathcal{C}_{\rm b}\). The latter removes only the interaction-mediated cyclic contribution, whereas full blocking of \(\mathcal{J}_R^{\rm ss}\) requires the cyclic and right-contact pumping contributions to cancel together. In the limit \(A_R=0\), the pumping numerator vanishes, \(\mathcal{B}_R=0\), and Eq.~\eqref{eq:dual_side_blocking_condition} reduces to the single-side-driven blocking condition.

\begin{figure*}[t]
\centering
\subfigure[]{
\includegraphics[width=0.32\textwidth]{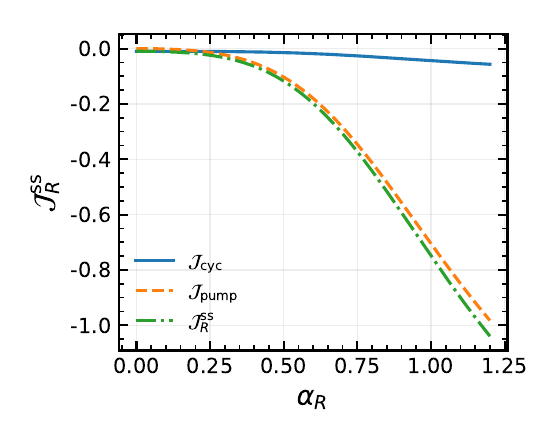}
\label{fig:fig_3a}
}
\hfill
\subfigure[]{
\includegraphics[width=0.32\textwidth]{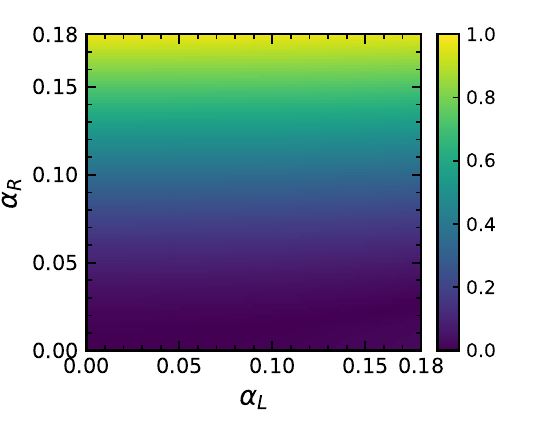}
\label{fig:fig_3b}
}
\hfill
\subfigure[]{
\includegraphics[width=0.32\textwidth]{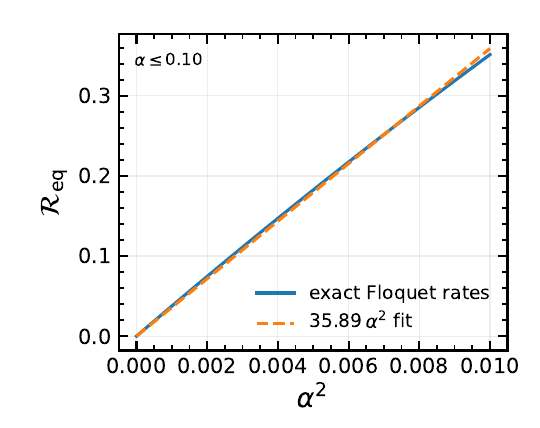}
\label{fig:fig_3c}
}
\caption{Numerical characterization of the dual-side-driven configuration.
(a) Exact forward right-bath current decomposed into the interaction-mediated cyclic contribution \(\mathcal{J}_{\rm cyc}\), the right-contact Floquet pumping contribution \(\mathcal{J}_{\rm pump}\), and their sum \(\mathcal{J}_{R}^{\rm ss}\), using Eq.~\eqref{eq:dual_side_factorized_current}. The right-drive strength \(\alpha_R\) is varied at fixed \(\alpha_L=0.65\).
(b) Rectification coefficient \(\mathcal{R}_{\rm dd}\) in the weak-drive flat-spectrum regime over \(0\leq\alpha_L,\alpha_R\leq0.18\), illustrating the unequal-amplitude scaling in Eq.~\eqref{eq:dual_side_rectification_unequal}.
(c) Equal-amplitude cut \(\alpha_L=\alpha_R=\alpha\), comparing the numerical Floquet result with the quadratic weak-drive prediction in Eq.~\eqref{eq:dual_side_equal_rectification}.
Parameters are \(\omega_0=6\), \(J=0.85\), \(\Omega=2.12\), \(T_h=4\), \(T_c=2\), and \(\gamma_L=\gamma_R=1\), with \(\hbar=k_B=1\).
}
\label{fig:fig3}
\end{figure*}

To obtain an analytical design rule, we expand the exact dual-side current in Eq.~\eqref{eq:dual_side_factorized_current} in the weak-drive flat-spectrum regime. We use the same symmetric resonant reference point as in the undriven and single-side analyses, namely \(\omega_{L,0}=\omega_{R,0}\equiv\omega_0\) and \(\kappa_L=\kappa_R\equiv\kappa\). The corresponding conditional transition frequencies are \(\nu_\sigma=\omega_0+\sigma\Omega_J\), and the weak-modulation parameters are \(\alpha_a\ll1\). Using the general sideband definition in Eq.~\eqref{eq:app_Pa}, the sinusoidal protocol in Eq.~\eqref{eq:dual_side_drive_protocol} gives \(P_a(q)=\mathrm J_q^2(\alpha_a)\). Its weak-drive expansion is
\begin{equation}
P_a(0)
=
1-\frac{\alpha_a^2}{2}
+
\mathcal{O}(\alpha_a^4),
\qquad
P_a(\pm1)
=
\frac{\alpha_a^2}{4}
+
\mathcal{O}(\alpha_a^4).
\label{eq:dual_side_weak_sideband_weights}
\end{equation}
Thus, only \(q=0,\pm1\) contributes through order \(\alpha_a^2\), and the leading drive-induced correction to the current is quadratic in the modulation amplitude. Accordingly, retaining terms through second order in $\alpha$ for Eq.~\eqref{eq:dual_side_factorized_current} gives
\begin{equation}
\mathcal{J}_{R,\rm flat}^{\rm ss}
=
\mathcal{J}_0
+
\alpha_L^2\mathcal{J}_{\mathrm L}^{(2)}
+
\alpha_R^2\mathcal{J}_{\mathrm R}^{(2)}
+
\mathcal{O}
\!\left(
\alpha_L^4,
\alpha_R^4,
\alpha_L^2\alpha_R^2
\right).
\label{eq:dual_side_weak_current}
\end{equation}
The zeroth-order term is the undriven symmetric-resonant current,
\begin{equation}
\mathcal{J}_0
=
2\hbar\Omega_J\kappa
\frac{\delta_0}{z_0},
\label{eq:dual_side_J0}
\end{equation}
and the two quadratic corrections are
\begin{align}
\mathcal{J}_{\mathrm L}^{(2)}
&=
2\hbar\Omega_J\kappa
\left[
\frac{\delta_L}{z_0}
-
\frac{\delta_0 z_L}{z_0^2}
\right],
\nonumber\\
\mathcal{J}_{\mathrm R}^{(2)}
&=
\hbar\kappa
\left[
2\Omega_J
\left(
\frac{\delta_R}{z_0}
-
\frac{\delta_0 z_R}{z_0^2}
\right)
+
\Omega
\frac{b_R}{z_0}
\right].
\label{eq:dual_side_quadratic_corrections}
\end{align}
The coefficients \(\delta_0\) and \(z_0\) determine the undriven normalized cycle current, while \(\delta_L,\delta_R\) and \(z_L,z_R\) describe the weak-drive corrections to the cycle imbalance and the steady-state normalization. Their explicit definitions, together with the pumping coefficient \(b_R\), are given in Appendix~\ref{app:dual_side_weak_drive_expansion}. In particular, Eq.~\eqref{eq:app_dual_weak_current_final} derives Eq.~\eqref{eq:dual_side_weak_current} directly from the exact right-bath current.

Equation~\eqref{eq:dual_side_weak_current} separates the two effects of dual-side modulation. Both drives modify the interaction-mediated circulation current through the imbalance and normalization corrections. However, only the right drive contributes the additional term proportional to \(\Omega b_R\), because the measured current is the right-bath heat current and the right bath exchanges sideband energy \(q\Omega\) when the right qubit is modulated. Thus, dual-side driving does not simply double the single-side correction; the second drive can either reinforce or compete with the rectifying contribution, depending on the signs and magnitudes of the circulation and pumping terms.

Having obtained the quadratic weak-drive corrections to the right-bath current in Eq.~\eqref{eq:dual_side_quadratic_corrections}, we now insert
the forward and reverse currents defined in Eq.~\eqref{eq:results_forward_reverse_currents} into the rectification coefficient in Eq.~\eqref{eq:results_rectification_coefficient}. In the
symmetric resonant reference case, the undriven contribution is reciprocal,
\begin{equation}
\mathcal{J}_0^B
=
-\mathcal{J}_0^F .
\label{eq:dual_side_undriven_reciprocity}
\end{equation}
Hence the zeroth-order term cancels from the numerator of $\mathcal{R}$, and the leading nonreciprocal contribution appears at quadratic order in the modulation amplitudes.
Retaining terms through this order gives
\begin{equation}
\mathcal{R}_{\rm dd}
=
\frac{
\left|
\alpha_L^2\mathcal{C}_L
+
\alpha_R^2\mathcal{C}_R
\right|
}{
|\mathcal{J}_0^F|
}
+
\mathcal{O}(\alpha^4).
\label{eq:dual_side_rectification_unequal}
\end{equation}
The coefficients appearing in Eq.~\eqref{eq:dual_side_rectification_unequal} collect the forward and reverse quadratic corrections generated by the left and right
modulations, respectively:
\begin{equation}
\mathcal{C}_L
=
\mathcal{J}_{\mathrm{L}}^{(2),F}
+
\mathcal{J}_{\mathrm{L}}^{(2),B},
\qquad
\mathcal{C}_R
=
\mathcal{J}_{\mathrm{R}}^{(2),F}
+
\mathcal{J}_{\mathrm{R}}^{(2),B}.
\label{eq:dual_side_rectification_coefficients}
\end{equation}
Equation~\eqref{eq:dual_side_rectification_unequal} therefore provides a perturbative design criterion for unequal-amplitude dual-side modulation.
If \(\mathcal{C}_L\) and \(\mathcal{C}_R\) have the same sign, the two modulations contribute constructively to the leading rectifying correction. If they have opposite signs, the two corrections compete, so increasing the second modulation can reduce the rectification generated by the first.

The equal-amplitude limit provides a useful symmetry check on the dual-side weak-drive correction. Setting \(\alpha_L=\alpha_R = \alpha\) in Eq.~\eqref{eq:dual_side_quadratic_corrections}, the right-bath current reduces to
\begin{equation}
\mathcal{J}_{R,\rm flat}^{\rm ss}
=
\mathcal{J}_0
+
\alpha^2\mathcal{J}_{\rm eq}^{(2)}
+
\mathcal{O}(\alpha^4),
\label{eq:dual_side_equal_current}
\end{equation}
where using the explicit left- and right-drive corrections in Eq.~\eqref{eq:dual_side_quadratic_corrections} gives
\begin{equation}
\mathcal{J}_{\rm eq}^{(2)}
=
\hbar\kappa
\left[
2\Omega_J
\left(
\frac{\delta_L+\delta_R}{z_0}
-
\frac{\delta_0(z_L+z_R)}{z_0^2}
\right)
+
\Omega
\frac{b_R}{z_0}
\right].
\label{eq:dual_side_equal_correction}
\end{equation}
The two terms in Eq.~\eqref{eq:dual_side_equal_correction} have different physical origins. The part proportional to \(2\Omega_J\) modifies the interaction-mediated cycle current through the corrections
\(\delta_L+\delta_R\) and \(z_L+z_R\). The part proportional to \(\Omega\), by contrast, is the right-contact Floquet pumping contribution associated with drive-assisted energy exchange at the
measured bath.

When the resonant reference system is left--right symmetric and the two modulations are identical, the cycle-current correction is reciprocal under \(T_L\leftrightarrow T_R\). Thus its forward and reverse contributions cancel in \(\mathcal{J}_F+\mathcal{J}_B\). The remaining contribution to the measured right-bath current asymmetry is then
\begin{equation}
\mathcal{J}_F+\mathcal{J}_B
=
\alpha^2\hbar\kappa\Omega
\frac{
b_R^F+b_R^B
}{
z_0^F
}
+
\mathcal{O}(\alpha^4).
\label{eq:dual_side_equal_nonreciprocal_current}
\end{equation}
Substituting this result into the rectification coefficient in Eq.~\eqref{eq:results_rectification_coefficient}, and using the leading-order current magnitude, yields
\begin{equation}
\mathcal{R}_{\rm eq}
=
\alpha^2
\frac{\Omega}{2\Omega_J}
\frac{
|b_R^F+b_R^B|
}{
|\delta_0^F|
}
+
\mathcal{O}(\alpha^4).
\label{eq:dual_side_equal_rectification}
\end{equation}
This limiting result shows that equal-amplitude dual-side driving does not simply double the single-side correction. In the fully symmetric configuration, the cycle-current correction remains reciprocal at quadratic order, while the residual contribution to the measured right-bath current arises from right-contact Floquet pumping. This distinction is important because, in a driven system, a nonzero contribution to \(\mathcal{J}_F+\mathcal{J}_B\) can originate from drive-assisted pumping rather than from a reinforced thermal-cycle rectification mechanism.

Figure~\ref{fig:fig3} summarizes the numerical behavior of the dual-side-driven configuration and verifies the analytical separation developed above. Panel~\ref{fig:fig_3a} evaluates the exact current decomposition in Eq.~\eqref{eq:dual_side_factorized_current} by varying the right-drive amplitude at fixed left drive. Unlike the single-side-driven case, the measured right contact is now Floquet dressed, so the right-bath current contains both the interaction-mediated cyclic contribution and the additional pumping term. The growth of the pumping contribution with \(\alpha_R\) illustrates why current blocking in the dual-side-driven case requires more than balancing the cyclic current alone. We then restrict to the weak-drive flat-spectrum regime in panels~\ref{fig:fig_3b} and \ref{fig:fig_3c}, where the perturbative results apply. Panel~\ref{fig:fig_3b} visualizes the unequal-amplitude scaling in Eq.~\eqref{eq:dual_side_rectification_unequal}, showing that \(\alpha_L^2\) and \(\alpha_R^2\) provide independent leading-order controls of \(\mathcal{R}_{\rm dd}\). The equal-amplitude cut in panel~\ref{fig:fig_3c} confirms the quadratic scaling predicted by Eq.~\eqref{eq:dual_side_equal_rectification}, while emphasizing that the residual nonreciprocal response is tied to right-contact Floquet pumping rather than a simple reinforcement of the single-side diode mechanism.

Our model admits a plausible superconducting-circuit realization based on two independently flux-tunable qubits, such as SQUID-based transmons. The qubits can be coupled through an engineered effective (ZZ) interaction, while separate ac-flux lines provide local modulation of their transition frequencies. Tunable cross-Kerr-type (ZZ) interactions have been demonstrated using flux-tunable couplers \cite{Collodo2020}, while strong capacitively engineered (ZZ) interactions have recently been reported in coupled transmons \cite{Riccardi2026}. Local ac-flux modulation has also been used to control qubit transition frequencies and generate sideband-assisted transitions \cite{Strand2013,Li2025EngineeredDissipation}. Independently heated mesoscopic reservoirs and photonic heat-current measurements based on electronic thermometry have been realized in superconducting quantum-transport experiments \cite{Senior2020}. More recently, two flux-tunable transmons, individual longitudinal flux lines, independently controlled thermal microwave reservoirs, and direct heat-current detection have been integrated within a single superconducting device \cite{Sundelin2026}. Collectively, these experiments demonstrate most of the hardware required by the proposed architecture, although the cited devices do not yet realize the exact combination of a dominant (ZZ) interaction, independent local thermal contacts, and local periodic Floquet modulation assumed in our model. A device-specific implementation would therefore need to verify that the effective (ZZ) coupling remains sufficiently stable during modulation, while leakage, residual exchange interactions, flux-line crosstalk, and reservoir cross-coupling are kept small relative to the intended dynamical scales. The reservoir spectral densities should also be sufficiently broad and slowly varying across the relevant conditional transitions and Floquet sidebands if the wide-band regime used in our analysis is to be approached experimentally. These considerations support the experimental plausibility of the proposed architecture, although a complete multilevel circuit analysis and parameter optimization lie beyond the scope of the present work.


\section{Conclusions}\label{sec:conclusions}

We have established asymmetric Floquet dressing as a mechanism for selective
heat-current suppression in a minimal quantum thermal diode formed by two
Ising-coupled qubits, and we have derived an exact criterion for complete
current blocking. Starting from a microscopic system--bath model, we derived a Floquet--LGKS
master equation that resolves the drive-assisted transition channels generated
by longitudinal periodic modulation.
The resonant undriven configuration with left--right symmetric bath couplings
provides the reciprocal benchmark under temperature reversal,
consistent with the known behavior of the static
model~\cite{PhysRevE.95.022128}.
In this static reference model, detuning can generate rectification, but
typically at the cost of a reduced transmitted heat
current~\cite{PhysRevE.95.022128}.
Driving only one qubit instead creates a Floquet-dressed contact opposite a
purely thermal contact, thereby breaking the reciprocal transport structure of
the symmetric undriven device.
This contact asymmetry provides an exact condition for blocking one
temperature-bias direction while the opposite direction remains conducting.
For the parameters considered numerically, one current branch approaches this
blocking condition while the other remains finite.

For the single-side-driven configuration, the steady-state current takes a
compact factorized form in which a positive kinetic prefactor multiplies a
competition factor that determines both the direction and the zeros of the
current.
This structure yields an exact analytical blocking condition and exposes the
sector-resolved imbalance responsible for directional heat transport.
In the weak sinusoidal-drive and flat-spectrum case, the rectification
coefficient receives its first nonvanishing contribution at quadratic order in
the drive amplitude.
Near the reciprocal reference point, this result shows that the leading
asymmetry arises from Floquet-sideband redistribution rather than from static
detuning.
The numerical results verify the selective suppression of one current branch
and confirm the predicted quadratic onset of rectification.
Together, these findings establish the operating principle of the
single-side-driven diode: Floquet modulation redistributes the spectral weights
of the Ising-resolved transition channels so that one bias direction approaches
current blocking without uniformly suppressing transport in the opposite
direction.

When both qubits are driven, the right-bath heat current separates into an
interaction-mediated cyclic contribution and a right-contact Floquet pumping
contribution. Consequently, balancing the cyclic population current alone is insufficient for
complete blocking, because the full current vanishes only when the cyclic and
pumping contributions cancel. The weak-drive expression further shows that the second modulation contributes
constructively when the two quadratic coefficients have the same sign, but
competes with the first modulation when their signs are opposite.
In the symmetric equal-amplitude configuration, the cycle-current correction
remains reciprocal at quadratic order, while the residual asymmetry of the
measured right-bath current originates from Floquet pumping.
Within the present model, dual-side driving is therefore not generically
superior to the single-side geometry, which provides the clearest realization
of the contact-asymmetry mechanism.
Superconducting circuits provide a plausible platform for testing this
mechanism, although integrating strong \(ZZ\) coupling, local Floquet
modulation, and independently controlled thermal reservoirs in a single device
remains experimentally challenging.
Taken together, these results establish contact-selective longitudinal Floquet
control as an analytical design principle and provide a controllable route to
strong heat-flow asymmetry in a minimal quantum thermal device.

\begin{acknowledgments}
M. Tahir Naseem acknowledges support from the National Center for Quantum Computing (NCQC), Pakistan.
\end{acknowledgments}

\appendix
\begin{widetext}

\section{Microscopic derivation of the Floquet--LGKS master equation}
\label{app:Floquet_LGKS_derivation}

This appendix derives the Floquet--LGKS master equation quoted in Eq.~\eqref{eq:main_Floquet_LGKS_IP} from the microscopic model introduced in Eq.~\eqref{eq:Htot}. We consider the general configuration in which both qubits are longitudinally and periodically driven with common period $\tau$, while each qubit remains weakly coupled to an independent thermal reservoir.

To set up the interaction-picture dynamics, we split the total Hamiltonian as
\begin{equation*}
H_{\mathrm{tot}}(t)
=
H_0(t)+\lambda_L H_{I,L}+\lambda_R H_{I,R},
\qquad
H_0(t)=H_S(t)+H_B.
\end{equation*}
Because the system Hamiltonian $H_S(t)$ in Eq.~\eqref{eq:HS_general} is diagonal in the computational basis at all times, it satisfies $ [H_S(t),H_S(s)]=0, \,\, \forall\, t,s.$
This property allows the free propagator to factorize into system and bath parts,
\begin{equation*}
U_0(t,0)=U_S(t,0)\otimes U_B(t,0),
\qquad
U_B(t,0)=e^{-\,\frac{i}{\hbar}H_B t}.
\end{equation*}
To evaluate the system propagator explicitly, we introduce the accumulated phases
\begin{equation*}
\theta_a(t)=\int_0^t \omega_a(s)\,ds
=
\omega_{a,0}t+\zeta_a(t),
\qquad
a\in\{L,R\},
\end{equation*}
where
\begin{equation*}
\zeta_a(t)=\int_0^t \delta\omega_a(s)\,ds,
\qquad
\zeta_a(t+\tau)=\zeta_a(t).
\end{equation*}
The periodicity of $\zeta_a(t)$ follows from the zero-mean condition on $\delta\omega_a(t)$ introduced in Eq.~\eqref{eq:zero_mean_drive}. The system propagator can then be written exactly as
\begin{equation*}
U_S(t,0)
=
e^{-\,\frac{i}{2}\theta_L(t)\sigma_z^{(L)}}
e^{-\,\frac{i}{2}\theta_R(t)\sigma_z^{(R)}}
e^{-\,\frac{i}{\hbar}J t\,\sigma_z^{(L)}\sigma_z^{(R)}}.
\end{equation*}
The interaction-picture system operators are defined by
\begin{equation*}
S_a(t)=U_S^\dagger(t,0)\,S_a\,U_S(t,0).
\end{equation*}
Using $S_a=\sigma_x^{(a)}=\sigma_+^{(a)}+\sigma_-^{(a)}$, together with the Pauli algebra and the Ising frequency scale $\Omega_J$ from Eq.~\eqref{eq:OmegaJ}, one obtains
\begin{align}
S_L(t)
&=
e^{+i\theta_L(t)}e^{+i\Omega_J t\,\sigma_z^{(R)}}\sigma_+^{(L)}
+
e^{-i\theta_L(t)}e^{-i\Omega_J t\,\sigma_z^{(R)}}\sigma_-^{(L)}
\label{eq:app_SLt}
\\
S_R(t)
&=
e^{+i\theta_R(t)}e^{+i\Omega_J t\,\sigma_z^{(L)}}\sigma_+^{(R)}
+
e^{-i\theta_R(t)}e^{-i\Omega_J t\,\sigma_z^{(L)}}\sigma_-^{(R)} .
\label{eq:app_SRt}
\end{align}
Each interaction-picture coupling operator therefore contains two distinct structures: a periodic phase generated by the external drive and a conditional phase generated by the Ising interaction. This separation is the starting point for the Floquet and sector decomposition developed below. The bath operators evolve in the interaction picture as
\begin{equation*}
B_a(t)
=
e^{+\frac{i}{\hbar}H_B^{(a)}t}B_a e^{-\,\frac{i}{\hbar}H_B^{(a)}t},
\end{equation*}
so that the interaction-picture interaction Hamiltonian takes the form
\begin{equation*}
H_I(t)
=
\lambda_L\,S_L(t)\otimes B_L(t)
+
\lambda_R\,S_R(t)\otimes B_R(t).
\end{equation*}
The full interaction-picture density operator then obeys
\begin{equation}
\frac{d}{dt}\rho_I(t)
=
-\frac{i}{\hbar}[H_I(t),\rho_I(t)].
\label{eq:app_Liouville}
\end{equation}
Iterating Eq.~\eqref{eq:app_Liouville} once and tracing over the bath degrees of freedom gives the exact reduced equation
\begin{equation*}
\frac{d}{dt}\rho_{S,I}(t)
=
-\frac{i}{\hbar}\operatorname{Tr}_B\!\bigl([H_I(t),\rho_I(0)]\bigr)
-\frac{1}{\hbar^2}\int_0^t ds\,
\operatorname{Tr}_B\!\bigl([H_I(t),[H_I(s),\rho_I(s)]]\bigr),
\end{equation*}
where $\rho_{S,I}(t)=\operatorname{Tr}_B\rho_I(t)$. For a factorized initial state $\rho_{\mathrm{tot}}(0)=\rho_S(0)\otimes\rho_B$, with $\rho_B=\rho_{B_L}\otimes\rho_{B_R}$, the first-order contribution vanishes because the bath operators have zero thermal mean,
\begin{equation*}
\operatorname{Tr}_{B_a}\!\bigl(B_a(t)\rho_{B_a}\bigr)=0.
\end{equation*}
Applying the Born approximation then yields
\begin{equation*}
\frac{d}{dt}\rho_{S,I}(t)
=
-\frac{1}{\hbar^2}\int_0^t ds\,
\operatorname{Tr}_B\!\bigl([H_I(t),[H_I(s),\rho_{S,I}(s)\otimes\rho_B]]\bigr).
\end{equation*}
Because the two reservoirs are statistically independent, mixed bath correlations vanish and the kernel separates into left and right contributions. Introducing the stationary bath correlation functions
\begin{equation*}
C_a(\tau)
=
\operatorname{Tr}_{B_a}\!\bigl(B_a(\tau)B_a(0)\rho_{B_a}\bigr),
\end{equation*}
one arrives at the Redfield equation
\begin{equation}
\frac{d}{dt}\rho_{S,I}(t)
=
-\sum_{a=L,R}\frac{\lambda_a^2}{\hbar^2}\int_0^t ds\,
\Big\{
C_a(t-s)\,[S_a(t),S_a(s)\rho_{S,I}(s)]
+
C_a^*(t-s)\,[\rho_{S,I}(s)S_a(s),S_a(t)]
\Big\}.
\label{eq:app_Redfield}
\end{equation}
We now resolve the explicit time dependence of the interaction-picture system operators into Floquet sidebands and Ising sectors. The periodic phases in Eqs.~\eqref{eq:app_SLt} and \eqref{eq:app_SRt} are expanded in Floquet harmonics through the Fourier coefficients
\begin{equation*}
\xi_a(n)
=
\frac{1}{\tau}\int_0^\tau e^{i\zeta_a(t)}e^{-in\Omega t}\,dt,
\end{equation*}
so that
\begin{equation*}
e^{+i\zeta_a(t)}
=
\sum_{n\in\mathbb Z}\xi_a(n)e^{in\Omega t},
\qquad
e^{-i\zeta_a(t)}
=
\sum_{n\in\mathbb Z}\xi_a^*(-n)e^{in\Omega t}.
\end{equation*}
The corresponding sideband weights are
\begin{equation}
P_a(n)=|\xi_a(n)|^2,
\qquad
\sum_{n\in\mathbb Z}P_a(n)=1.
\label{eq:app_Pa}
\end{equation}
To resolve the conditional sectors generated by the Ising interaction, we use the projectors
\begin{equation*}
\Pi_\pm^{(L)}=\frac{1}{2}\bigl(\mathbbm{1}\pm\sigma_z^{(L)}\bigr),
\qquad
\Pi_\pm^{(R)}=\frac{1}{2}\bigl(\mathbbm{1}\pm\sigma_z^{(R)}\bigr),
\end{equation*}
which give
\begin{equation*}
e^{\pm i\Omega_J t\,\sigma_z^{(R)}}
=
\Pi_+^{(R)}e^{\pm i\Omega_J t}
+
\Pi_-^{(R)}e^{\mp i\Omega_J t},
\qquad
e^{\pm i\Omega_J t\,\sigma_z^{(L)}}
=
\Pi_+^{(L)}e^{\pm i\Omega_J t}
+
\Pi_-^{(L)}e^{\mp i\Omega_J t}.
\end{equation*}
Substituting these decompositions into Eqs.~\eqref{eq:app_SLt} and \eqref{eq:app_SRt}, the interaction-picture operators take the channel-resolved form
\begin{align*}
S_L(t)
&=
\sum_{q\in\mathbb Z}\sum_{\sigma=\pm}
\Big[
S_{L;q,\sigma}^{(+)}e^{+i\nu_{L,q}^{(\sigma)}t}
+
S_{L;q,\sigma}^{(-)}e^{-i\nu_{L,q}^{(\sigma)}t}
\Big]
\\
S_R(t)
&=
\sum_{p\in\mathbb Z}\sum_{\sigma=\pm}
\Big[
S_{R;p,\sigma}^{(+)}e^{+i\nu_{R,p}^{(\sigma)}t}
+
S_{R;p,\sigma}^{(-)}e^{-i\nu_{R,p}^{(\sigma)}t}
\Big] .
\end{align*}
with Bohr--Floquet frequencies
\begin{equation*}
\nu_{L,q}^{(\sigma)}
=
\omega_{L,0}+q\Omega+\sigma\Omega_J,
\qquad
\nu_{R,p}^{(\sigma)}
=
\omega_{R,0}+p\Omega+\sigma\Omega_J,
\end{equation*}
and channel operators
\begin{alignat}{2}
S_{L;q,\sigma}^{(+)}
&=
\xi_L(q)\Pi_\sigma^{(R)}\sigma_+^{(L)},
\qquad&
S_{L;q,\sigma}^{(-)}
&=
\xi_L^*(q)\Pi_\sigma^{(R)}\sigma_-^{(L)}
\label{eq:app_SL_blocks}
\\
S_{R;p,\sigma}^{(+)}
&=
\xi_R(p)\Pi_\sigma^{(L)}\sigma_+^{(R)},
\qquad&
S_{R;p,\sigma}^{(-)}
&=
\xi_R^*(p)\Pi_\sigma^{(L)}\sigma_-^{(R)} .
\label{eq:app_SR_blocks}
\end{alignat}
Thus, each bath-induced transition is resolved simultaneously into a Floquet ladder of sidebands and into the two conditional sectors generated by the Ising interaction.

Substituting the channel-resolved operators given in Eqs. \eqref{eq:app_SL_blocks}-\eqref{eq:app_SR_blocks} into the Redfield equation \eqref{eq:app_Redfield} produces terms oscillating at sums and differences of Bohr--Floquet frequencies. To obtain a time-local generator, we next apply the Markov approximation by replacing $\rho_{S,I}(s)$ with $\rho_{S,I}(t)$ inside the memory kernel and extending the upper limit of integration to infinity. This yields
\begin{equation}
\frac{d}{dt}\rho_{S,I}(t)
=
-\sum_{a=L,R}\frac{\lambda_a^2}{\hbar^2}\int_0^\infty d\tau\,
\Big\{
C_a(\tau)\,[S_a(t),S_a(t-\tau)\rho_{S,I}(t)]
+
C_a^*(\tau)\,[\rho_{S,I}(t)S_a(t-\tau),S_a(t)]
\Big\}.
\label{eq:app_MarkovRedfield}
\end{equation}
The required half-sided Fourier transforms of the bath correlation functions are
\begin{align*}
\int_0^\infty d\tau\,C_a(\tau)e^{i\omega\tau}
&=
\frac{1}{2}G_a(\omega)+i\Lambda_a(\omega),
\end{align*}
where $G_a(\omega)$ is the bath spectral function and $\Lambda_a(\omega)$ is the Lamb-shift kernel. Thermal equilibrium further implies the KMS relation
\begin{equation*}
G_a(-\omega)=e^{-\beta_a\hbar\omega}G_a(\omega).
\end{equation*}

At this stage, the generator \eqref{eq:app_MarkovRedfield} still contains explicit oscillatory factors coupling distinct Bohr--Floquet channels. These are removed by the Floquet--secular approximation, which discards rapidly oscillating contributions and retains only frequency-diagonal terms. In the generic nondegenerate case, this leaves channels labeled by $(q,\sigma)$ for the left bath and by $(p,\sigma)$ for the right bath. The frequency-diagonal terms selected by the Floquet--secular approximation define the jump operators
\begin{align}
L_{q,\sigma,\downarrow}^{(L)}
&=
\Pi_\sigma^{(R)}\sigma_-^{(L)},
\qquad
L_{q,\sigma,\uparrow}^{(L)}
=
\Pi_\sigma^{(R)}\sigma_+^{(L)} ,
\label{eq:app_JumpsL}
\\
L_{p,\sigma,\downarrow}^{(R)}
&=
\Pi_\sigma^{(L)}\sigma_-^{(R)},
\qquad
L_{p,\sigma,\uparrow}^{(R)}
=
\Pi_\sigma^{(L)}\sigma_+^{(R)} .
\label{eq:app_JumpsR}
\end{align}
These operators describe downward and upward qubit flips resolved by the conditional sector of the opposite qubit. Their operator structure is independent of the sideband labels $q$ and $p$; the sideband dependence enters through the corresponding transition rates. For the left and right baths, the rates are
\begin{align}
\Gamma_{\downarrow}^{(L)}(q,\sigma)
&=
\frac{\lambda_L^2}{\hbar^2}P_L(q)G_L\!\bigl(\nu_{L,q}^{(\sigma)}\bigr),
\qquad
\Gamma_{\uparrow}^{(L)}(q,\sigma)
=
\frac{\lambda_L^2}{\hbar^2}P_L(q)G_L\!\bigl(-\nu_{L,q}^{(\sigma)}\bigr),
\label{eq:app_RatesL}
\\
\Gamma_{\downarrow}^{(R)}(p,\sigma)
&=
\frac{\lambda_R^2}{\hbar^2}P_R(p)G_R\!\bigl(\nu_{R,p}^{(\sigma)}\bigr),
\qquad
\Gamma_{\uparrow}^{(R)}(p,\sigma)
=
\frac{\lambda_R^2}{\hbar^2}P_R(p)G_R\!\bigl(-\nu_{R,p}^{(\sigma)}\bigr).
\label{eq:app_RatesR}
\end{align}
Each channel therefore carries two distinct pieces of information: the sideband weight $P_a$, which encodes the effect of the periodic modulation, and the spectral factor $G_a(\pm\nu)$, which evaluates the bath response at the corresponding Bohr--Floquet frequency. By construction, these rates satisfy detailed balance channel by channel through the KMS relation.
The coherent principal-value contributions of the bath response generate the Lamb-shift Hamiltonian
\begin{align}
H_{\mathrm{LS}}
&=
\frac{\lambda_L^2}{\hbar}
\sum_{q\in\mathbb Z}\sum_{\sigma=\pm}
P_L(q)
\Big[
\Lambda_L\!\bigl(\nu_{L,q}^{(\sigma)}\bigr)\Pi_\sigma^{(R)}\sigma_+^{(L)}\sigma_-^{(L)}
+
\Lambda_L\!\bigl(-\nu_{L,q}^{(\sigma)}\bigr)\Pi_\sigma^{(R)}\sigma_-^{(L)}\sigma_+^{(L)}
\Big]
\nonumber\\
&\quad
+
\frac{\lambda_R^2}{\hbar}
\sum_{p\in\mathbb Z}\sum_{\sigma=\pm}
P_R(p)
\Big[
\Lambda_R\!\bigl(\nu_{R,p}^{(\sigma)}\bigr)\Pi_\sigma^{(L)}\sigma_+^{(R)}\sigma_-^{(R)}
+
\Lambda_R\!\bigl(-\nu_{R,p}^{(\sigma)}\bigr)\Pi_\sigma^{(L)}\sigma_-^{(R)}\sigma_+^{(R)}
\Big].
\label{eq:app_HLS}
\end{align}
Combining this coherent contribution Eq. \eqref{eq:app_HLS} with the dissipative channels Eqs.~\eqref{eq:app_RatesL} and \eqref{eq:app_RatesR} yields the interaction-picture Floquet--LGKS equation
\begin{align}
\frac{d}{dt}\rho_{S,I}(t)
&=
-\frac{i}{\hbar}[H_{\mathrm{LS}},\rho_{S,I}(t)]
\nonumber\\
&\quad
+\sum_{q\in\mathbb Z}\sum_{\sigma=\pm}
\Big[
\Gamma_{\downarrow}^{(L)}(q,\sigma)\mathcal D\!\bigl[L_{q,\sigma,\downarrow}^{(L)}\bigr]\rho_{S,I}(t)
+
\Gamma_{\uparrow}^{(L)}(q,\sigma)\mathcal D\!\bigl[L_{q,\sigma,\uparrow}^{(L)}\bigr]\rho_{S,I}(t)
\Big]
\nonumber\\
&\quad
+\sum_{p\in\mathbb Z}\sum_{\sigma=\pm}
\Big[
\Gamma_{\downarrow}^{(R)}(p,\sigma)\mathcal D\!\bigl[L_{p,\sigma,\downarrow}^{(R)}\bigr]\rho_{S,I}(t)
+
\Gamma_{\uparrow}^{(R)}(p,\sigma)\mathcal D\!\bigl[L_{p,\sigma,\uparrow}^{(R)}\bigr]\rho_{S,I}(t)
\Big].
\label{eq:app_LGKS_IP}
\end{align}
The single-side-driven and undriven limits follow immediately by setting one or both modulation functions in Eq.~\eqref{eq:freq_decomposition} to zero, in which case the corresponding Floquet ladders collapse to their central components.

\section{Static dynamics and normalization factor for the undriven benchmark}
\label{app:undriven_normalization_factor_appendix}

For completeness, we collect here the static dynamical ingredients used in Sec.~\ref{subsec:undriven_reference_case}. In the absence of longitudinal modulation, the Floquet ladders collapse to their central components, $q=0$, and the reduced dynamics becomes the sector-resolved static LGKS equation \cite{PhysRevE.95.022128}
\begin{equation}
\begin{aligned}
\dot{\rho}
=
-\frac{i}{\hbar}[H_S,\rho] +\sum_{\sigma=\pm}\sum_{a=L,R}
\Big[
\Gamma_{\downarrow}^{(a,\sigma)}
\mathcal D\!\bigl[L_{\downarrow}^{(a,\sigma)}\bigr]\rho
+
\Gamma_{\uparrow}^{(a,\sigma)}
\mathcal D\!\bigl[L_{\uparrow}^{(a,\sigma)}\bigr]\rho
\Big],
\end{aligned}
\label{eq:undriven_static_ME}
\end{equation}
The jump operators in Eq.~\eqref{eq:undriven_static_ME} resolve the transition of one qubit according to the state of the other qubit. They are
\begin{equation*}
\begin{aligned}
L_{\downarrow}^{(L,\sigma)}
&=
\Pi_{\sigma}^{(R)}\sigma_-^{(L)},
&
L_{\uparrow}^{(L,\sigma)}
&=
\Pi_{\sigma}^{(R)}\sigma_+^{(L)},
&
L_{\downarrow}^{(R,\sigma)}
&=
\Pi_{\sigma}^{(L)}\sigma_-^{(R)},
&
L_{\uparrow}^{(R,\sigma)}
&=
\Pi_{\sigma}^{(L)}\sigma_+^{(R)} .
\end{aligned}
\end{equation*}
Thus each bath-induced transition is labelled by the bath index $a=L,R$ and by the conditional sector $\sigma=\pm$. For the flat bare spectral density used in the transport analysis, we take
\begin{equation}
G_a(\omega)=\gamma_a\bigl[\bar n_a(\omega)+1\bigr]\quad(\omega>0),
\qquad\qquad
G_a(\omega)=\gamma_a\,\bar n_a\bigl(|\omega|\bigr)\quad(\omega<0)
\label{eq:undriven_flat_spectrum}
\end{equation}
where
\begin{equation*}
\bar n_a(\omega)
=
\frac{1}{e^{\beta_a\hbar\omega}-1}
\end{equation*}
is the Bose occupation factor of bath $a$. The corresponding relaxation and excitation rates are evaluated at the conditional transition frequencies $\nu_a^{(\sigma)}$:
\begin{equation}
\Gamma_{\downarrow}^{(a,\sigma)}
=
\frac{\lambda_a^2}{\hbar^2}\,
G_a\!\bigl(\nu_a^{(\sigma)}\bigr),
\qquad
\Gamma_{\uparrow}^{(a,\sigma)}
=
\frac{\lambda_a^2}{\hbar^2}\,
G_a\!\bigl(-\nu_a^{(\sigma)}\bigr).
\label{eq:undriven_rate_definitions}
\end{equation}
It follows directly from Eq.~\eqref{eq:undriven_flat_spectrum} that these rates satisfy the detailed-balance relation
\begin{equation}
\Gamma_{\uparrow}^{(a,\sigma)}
=
r_a^{(\sigma)}\,\Gamma_{\downarrow}^{(a,\sigma)},
\qquad
r_a^{(\sigma)}
=
e^{-\beta_a\hbar\nu_a^{(\sigma)}} .
\label{eq:undriven_DB_rates}
\end{equation}
Solving the steady-state population equations generated by Eq.~\eqref{eq:undriven_static_ME} gives the normalization factor $\mathcal Z$ that appears in the undriven current expression, Eq.~\eqref{eq:undriven_right_current_final}. In terms of the sector-resolved relaxation and excitation rates, it is
\begin{equation}
\begin{aligned}
\mathcal Z
&=
\Gamma_{\downarrow}^{(L,+)}\Gamma_{\downarrow}^{(L,-)}\Gamma_{\downarrow}^{(R,-)}
+\Gamma_{\downarrow}^{(L,+)}\Gamma_{\downarrow}^{(L,-)}\Gamma_{\uparrow}^{(R,-)}
+\Gamma_{\downarrow}^{(L,+)}\Gamma_{\uparrow}^{(L,-)}\Gamma_{\uparrow}^{(R,+)}
+\Gamma_{\downarrow}^{(L,+)}\Gamma_{\uparrow}^{(L,-)}\Gamma_{\downarrow}^{(R,-)}
\\[2pt]
&\quad
+\Gamma_{\downarrow}^{(L,+)}\Gamma_{\uparrow}^{(R,+)}\Gamma_{\downarrow}^{(R,-)}
+\Gamma_{\downarrow}^{(L,+)}\Gamma_{\uparrow}^{(R,+)}\Gamma_{\uparrow}^{(R,-)}
+\Gamma_{\uparrow}^{(L,+)}\Gamma_{\downarrow}^{(L,-)}\Gamma_{\downarrow}^{(R,+)}
+\Gamma_{\uparrow}^{(L,+)}\Gamma_{\downarrow}^{(L,-)}\Gamma_{\uparrow}^{(R,-)}
\\[2pt]
&\quad
+\Gamma_{\uparrow}^{(L,+)}\Gamma_{\uparrow}^{(L,-)}\Gamma_{\downarrow}^{(R,+)}
+\Gamma_{\uparrow}^{(L,+)}\Gamma_{\uparrow}^{(L,-)}\Gamma_{\uparrow}^{(R,+)}
+\Gamma_{\uparrow}^{(L,+)}\Gamma_{\downarrow}^{(R,+)}\Gamma_{\uparrow}^{(R,-)}
+\Gamma_{\uparrow}^{(L,+)}\Gamma_{\uparrow}^{(R,+)}\Gamma_{\uparrow}^{(R,-)}
\\[2pt]
&\quad
+\Gamma_{\downarrow}^{(L,-)}\Gamma_{\downarrow}^{(R,+)}\Gamma_{\downarrow}^{(R,-)}
+\Gamma_{\downarrow}^{(L,-)}\Gamma_{\downarrow}^{(R,+)}\Gamma_{\uparrow}^{(R,-)}
+\Gamma_{\uparrow}^{(L,-)}\Gamma_{\downarrow}^{(R,+)}\Gamma_{\downarrow}^{(R,-)}
+\Gamma_{\uparrow}^{(L,-)}\Gamma_{\uparrow}^{(R,+)}\Gamma_{\downarrow}^{(R,-)} .
\end{aligned}
\label{eq:app_undriven_normalization_factor}
\end{equation}
Since all relaxation and excitation rates are non-negative, $\mathcal Z>0$. Therefore, this normalization factor does not affect the sign of the undriven current; the sign is determined entirely by the detailed-balance combination appearing in Eq.~\eqref{eq:undriven_current_bracket_symmetric}.

\section{Single-side driven transport: exact current, blocking conditions, and weak-drive expansion}
\label{app:single_side_transport}

This appendix derives the analytical results used in Sec.~\ref{subsec:single_side_driving}. We first obtain the steady-state population dynamics and the exact right-bath current for the single-side-driven configuration. We then derive the general blocking condition directly from the factorized current, without specifying a driving waveform or a spectral-density model. Finally, we specialize to the sinusoidal flat-spectrum realization and perform the weak-drive expansion leading to the quadratic scaling law in Eq.~\eqref{eq:single_side_weak_drive_scaling}.

\subsection{Population dynamics and exact steady-state current}
\label{appsubsec:single_side_current_derivation}

We begin with the single-side-driven configuration defined in Eq.~\eqref{eq:single_side_drive_specialization}, in which only the left qubit is longitudinally modulated. In this limit, the left bath remains Floquet and sector resolved, whereas the right bath reduces to two static sector channels. Since the Lamb-shift contribution does not affect the population dynamics relevant for the steady-state current, we omit it in what follows. The interaction-picture generator then reduces to
\begin{equation}
\dot{\rho}_{S,I}(t)
=
\mathcal L_L \rho_{S,I}(t)
+
\mathcal L_R \rho_{S,I}(t),
\label{eq:app_single_side_generator}
\end{equation}
with
\begin{align*}
\mathcal L_L \rho
&=
\sum_{q\in\mathbb Z}\sum_{\sigma=\pm}
\Big[
\Gamma_{\downarrow}^{(L)}(q,\sigma)\,
\mathcal D\!\bigl[\Pi_\sigma^{(R)}\sigma_-^{(L)}\bigr]\rho
+
\Gamma_{\uparrow}^{(L)}(q,\sigma)\,
\mathcal D\!\bigl[\Pi_\sigma^{(R)}\sigma_+^{(L)}\bigr]\rho
\Big],
\\
\mathcal L_R \rho
&=
\sum_{\sigma=\pm}
\Big[
\Gamma_{\downarrow}^{(R)}(\sigma)\,
\mathcal D\!\bigl[\Pi_\sigma^{(L)}\sigma_-^{(R)}\bigr]\rho
+
\Gamma_{\uparrow}^{(R)}(\sigma)\,
\mathcal D\!\bigl[\Pi_\sigma^{(L)}\sigma_+^{(R)}\bigr]\rho
\Big].
\end{align*}
For compactness, we introduce the effective rates
\begin{equation}
a_\sigma
=
\sum_q \Gamma_{\downarrow}^{(L)}(q,\sigma),
\qquad
b_\sigma
=
\sum_q \Gamma_{\uparrow}^{(L)}(q,\sigma),
\qquad
c_\sigma
=
\Gamma_{\downarrow}^{(R)}(\sigma),
\qquad
d_\sigma
=
\Gamma_{\uparrow}^{(R)}(\sigma),
\label{eq:app_single_side_abcd}
\end{equation}
with $\sigma=\pm$. We work in the ordered computational basis $\{\ket{ee},\ket{eg},\ket{ge},\ket{gg}\}$, where the first entry refers to the left qubit and the second to the right qubit. In this basis, the left bath drives the transitions $\ket{ee}\leftrightarrow\ket{ge}$ and $\ket{eg}\leftrightarrow\ket{gg}$, whereas the right bath drives $\ket{ee}\leftrightarrow\ket{eg}$ and $\ket{ge}\leftrightarrow\ket{gg}$. Because each jump operator connects only one pair of basis states, the population sector is dynamically closed and decouples from the coherences. Accordingly, Eq.~\eqref{eq:app_single_side_generator} reduces to the closed rate equations
\begin{equation}
\begin{alignedat}{2}
\dot p_{ee}
&=
-(a_+ + c_+)p_{ee}
+ b_+ p_{ge}
+ d_+ p_{eg},
\qquad&
\dot p_{eg}
&=
c_+ p_{ee}
-(a_- + d_+)p_{eg}
+ b_- p_{gg},
\\
\dot p_{ge}
&=
a_+ p_{ee}
-(b_+ + c_-)p_{ge}
+ d_- p_{gg},
\qquad&
\dot p_{gg}
&=
a_- p_{eg}
+ c_- p_{ge}
-(b_- + d_-)p_{gg}.
\end{alignedat}
\label{eq:app_single_side_pop_eqs}
\end{equation}
Solving Eq.~\eqref{eq:app_single_side_pop_eqs} at steady state gives
\begin{equation}
p_{ee}^{\rm ss}
=
\frac{W_{ee}}{\mathcal Z_{\mathrm{sd}}},
\qquad
p_{eg}^{\rm ss}
=
\frac{W_{eg}}{\mathcal Z_{\mathrm{sd}}},
\qquad
p_{ge}^{\rm ss}
=
\frac{W_{ge}}{\mathcal Z_{\mathrm{sd}}},
\qquad
p_{gg}^{\rm ss}
=
\frac{W_{gg}}{\mathcal Z_{\mathrm{sd}}},
\label{eq:app_single_side_ss_pops}
\end{equation}
where
\begin{equation*}
\mathcal Z_{\mathrm{sd}}
=
W_{ee}+W_{eg}+W_{ge}+W_{gg},
\end{equation*}
and
\begin{equation*}
\begin{alignedat}{2}
W_{ee}
&=
a_- b_+ d_-
+ b_+ b_- d_+
+ b_+ d_+ d_-
+ b_- c_- d_+,
\qquad&
W_{eg}
&=
a_+ b_- c_-
+ b_+ b_- c_+
+ b_+ c_+ d_-
+ b_- c_+ c_-,
\\
W_{ge}
&=
a_+ a_- d_-
+ a_+ b_- d_+
+ a_+ d_+ d_-
+ a_- c_+ d_-,
\qquad&
W_{gg}
&=
a_+ a_- c_-
+ a_+ c_- d_+
+ a_- b_+ c_+
+ a_- c_+ c_- .
\end{alignedat}
\end{equation*}
By construction, $\mathcal Z_{\mathrm{sd}}>0$ for positive rates, so the steady state is properly normalized.

We now evaluate the right-bath steady-state heat current. With the convention that $\mathcal J_R^{\rm ss}>0$ denotes net energy flow from the right bath into the system, the two right-bath transition energies are
\begin{equation}
\hbar\nu_R^{(+)}
=
\hbar\omega_R+2J,
\qquad
\hbar\nu_R^{(-)}
=
\hbar\omega_R-2J.
\label{eq:app_single_side_right_freqs}
\end{equation}
Using Eq.~\eqref{eq:results_heat_current_undriven}, the right-bath current is
\begin{equation}
\mathcal J_R^{\rm ss}
=
\hbar\nu_R^{(+)}
\Big(
d_+ p_{eg}^{\rm ss}
-
c_+ p_{ee}^{\rm ss}
\Big)
+
\hbar\nu_R^{(-)}
\Big(
d_- p_{gg}^{\rm ss}
-
c_- p_{ge}^{\rm ss}
\Big).
\label{eq:app_single_side_JR_pop}
\end{equation}
Substituting Eq.~\eqref{eq:app_single_side_ss_pops} into Eq.~\eqref{eq:app_single_side_JR_pop} and using Eq.~\eqref{eq:app_single_side_right_freqs}, the current reduces to
\begin{equation}
\mathcal J_R^{\rm ss}
=
\frac{2\hbar\Omega_J}{\mathcal Z_{\mathrm{sd}}}
\Big(
a_+b_-c_-d_+
-
a_-b_+c_+d_-
\Big).
\label{eq:app_single_side_JR_compact}
\end{equation}
To expose the physical structure of Eq.~\eqref{eq:app_single_side_JR_compact}, we introduce the effective Floquet ratios
\begin{equation}
r_\sigma^{(L)}
=
\frac{b_\sigma}{a_\sigma}.
\label{eq:app_single_side_rsigma}
\end{equation}
Because the right bath is undriven, its rates satisfy the equilibrium detailed-balance relation
\begin{equation}
d_\sigma
=
e^{-\beta_R\hbar\nu_R^{(\sigma)}}c_\sigma.
\label{eq:app_single_side_right_DB}
\end{equation}
Substituting Eqs.~\eqref{eq:app_single_side_rsigma} and \eqref{eq:app_single_side_right_DB} into Eq.~\eqref{eq:app_single_side_JR_compact} gives
\begin{equation}
\mathcal J_R^{\rm ss}(T_L,T_R)
=
2\hbar\Omega_J\,
\mathcal K(T_L,T_R)\,
\Xi(T_L,T_R),
\label{eq:app_single_side_factorized}
\end{equation}
with
\begin{equation}
\mathcal K(T_L,T_R)
=
\frac{
a_+(T_L)a_-(T_L)c_+(T_R)c_-(T_R)
}{
\mathcal Z_{\mathrm{sd}}(T_L,T_R)
},
\label{eq:app_single_side_K}
\end{equation}
and
\begin{equation*}
\Xi(T_L,T_R)
=
r_-^{(L)}(T_L)e^{-\beta_R\hbar\nu_R^{(+)}}
-
r_+^{(L)}(T_L)e^{-\beta_R\hbar\nu_R^{(-)}}.
\end{equation*}
For finite couplings and nonzero transport rates, $\mathcal K(T_L,T_R)>0$. Therefore, the sign of the current is controlled entirely by the competition bracket $\Xi(T_L,T_R)$.

\subsection{Sinusoidal drive and flat-spectrum specialization}
\label{appsubsec:single_side_flat_spectrum}

We now specialize the general rates to the flat-spectrum model in
Eq.~\eqref{eq:undriven_flat_spectrum}, used in the analytical weak-drive
expansion and in Fig.~\ref{fig:fig2}. We assume
$\omega_R>|\Omega_J|$, so that both transition frequencies of the
undriven right contact lie on the positive-frequency branch. By contrast,
the driven-left Bohr--Floquet frequencies defined in
Eq.~\eqref{eq:single_side_nuLq} may have either sign. The left-contact
rates must therefore be evaluated using the appropriate positive- or
negative-frequency branch of Eq.~\eqref{eq:undriven_flat_spectrum}.
Exact zero-frequency channels, $\nu_{L,q}^{(\sigma)}=0$, are excluded. It is useful to define
\begin{equation}
\kappa_L
=
\frac{\lambda_L^2\gamma_L}{\hbar^2},
\qquad
\kappa_R
=
\frac{\lambda_R^2\gamma_R}{\hbar^2}.
\label{eq:app_single_side_kappas}
\end{equation}
For the sinusoidal modulation considered here, the sideband weights are
given by Eq.~\eqref{eq:single_side_sideband_weights}. Applying the
general rate definition in Eq.~\eqref{eq:app_RatesL} to the two
frequency branches of the flat-spectrum model gives
\begin{align*}
\Gamma_{\downarrow}^{(L)}(q,\sigma)
&=
\kappa_L P(q)
\begin{cases}
\bar n_L\!\bigl(\nu_{L,q}^{(\sigma)}\bigr)+1,
&
\nu_{L,q}^{(\sigma)}>0,
\\[1mm]
\bar n_L\!\bigl(\lvert\nu_{L,q}^{(\sigma)}\rvert\bigr),
&
\nu_{L,q}^{(\sigma)}<0,
\end{cases}
&
\Gamma_{\uparrow}^{(L)}(q,\sigma)
&=
\kappa_L P(q)
\begin{cases}
\bar n_L\!\bigl(\nu_{L,q}^{(\sigma)}\bigr),
&
\nu_{L,q}^{(\sigma)}>0,
\\[1mm]
\bar n_L\!\bigl(\lvert\nu_{L,q}^{(\sigma)}\rvert\bigr)+1,
&
\nu_{L,q}^{(\sigma)}<0.
\end{cases}
\end{align*}
To express the corresponding total rates compactly, define the effective
Floquet occupation and the positive- and negative-frequency sideband
weights as
\begin{equation}
\begin{aligned}
N_\sigma(T_L)
&=
\sum_q
P(q)\,
\bar n_L\!\bigl(\lvert\nu_{L,q}^{(\sigma)}\rvert\bigr),
\\
\mathcal W_\sigma^{>}
&=
\sum_{\{q:\,\nu_{L,q}^{(\sigma)}>0\}}P(q),
\qquad
\mathcal W_\sigma^{<}
=
\sum_{\{q:\,\nu_{L,q}^{(\sigma)}<0\}}P(q).
\end{aligned}
\label{eq:app_single_side_Nsigma}
\end{equation}
Because zero-frequency channels have been excluded, normalization of the
Floquet weights implies
$\mathcal W_\sigma^{>}+\mathcal W_\sigma^{<}=1$. The total left-contact
rates defined in Eq.~\eqref{eq:app_single_side_abcd} are therefore
\begin{equation}
a_\sigma
=
\kappa_L
\bigl[
N_\sigma(T_L)+\mathcal W_\sigma^{>}
\bigr],
\qquad
b_\sigma
=
\kappa_L
\bigl[
N_\sigma(T_L)+\mathcal W_\sigma^{<}
\bigr].
\label{eq:app_single_side_flat_ab}
\end{equation}
The corresponding driven-left-contact ratio is
\begin{equation}
r_\sigma^{(L)}(T_L)
=
\frac{
N_\sigma(T_L)+\mathcal W_\sigma^{<}
}{
N_\sigma(T_L)+\mathcal W_\sigma^{>}
}.
\label{eq:app_single_side_flat_rsigma}
\end{equation}
For the undriven right contact, we write
$m_\sigma(T_R)=\bar n_R\!\bigl(\nu_R^{(\sigma)}\bigr)$. Its transition
rates are
\begin{equation}
c_\sigma
=
\kappa_R
\bigl[1+m_\sigma(T_R)\bigr],
\qquad
d_\sigma
=
\kappa_R m_\sigma(T_R).
\label{eq:app_single_side_flat_cd}
\end{equation}
Substituting these rates into the compact current in
Eq.~\eqref{eq:app_single_side_JR_compact} gives
\begin{equation}
\mathcal J_R^{\rm ss}(T_L,T_R)
=
2\hbar\Omega_J\,
\mathcal K_{\rm flat}(T_L,T_R)\,
\Xi_{\rm flat}(T_L,T_R),
\label{eq:app_single_side_flat_factorized}
\end{equation}
where
\begin{equation}
\mathcal K_{\rm flat}(T_L,T_R)
=
\frac{
\kappa_L^2\kappa_R^2
\bigl(N_++\mathcal W_+^{>}\bigr)
\bigl(N_-+\mathcal W_-^{>}\bigr)
\bigl(1+m_+\bigr)
\bigl(1+m_-\bigr)
}{
\mathcal Z_{\mathrm{sd}}(T_L,T_R)
},
\label{eq:app_single_side_Kflat}
\end{equation}
and
\begin{equation}
\Xi_{\rm flat}(T_L,T_R)
=
r_-^{(L)}(T_L)
e^{-\beta_R\hbar\nu_R^{(+)}}
-
r_+^{(L)}(T_L)
e^{-\beta_R\hbar\nu_R^{(-)}}.
\label{eq:app_single_side_Xiflat}
\end{equation}
Equations~\eqref{eq:app_single_side_flat_factorized}--%
\eqref{eq:app_single_side_Xiflat} constitute the signed-frequency
flat-spectrum realization of the general factorized current in
Eq.~\eqref{eq:app_single_side_factorized}.

In the weak-drive expansion through $\mathcal O(\alpha^2)$, only the
carrier and first sidebands contribute. If
$\omega_{L,0}>\Omega+|\Omega_J|$, all channels retained at this order
have positive frequencies. Consequently,
\begin{align*}
\mathcal W_\sigma^{>}
&=
1+\mathcal O(\alpha^4),
&
\mathcal W_\sigma^{<}
&=
\mathcal O(\alpha^4),
\\
a_\sigma
&=
\kappa_L
\bigl[1+N_\sigma(T_L)\bigr]
+\mathcal O(\alpha^4),
&
b_\sigma
&=
\kappa_L N_\sigma(T_L)
+\mathcal O(\alpha^4),
&
r_\sigma^{(L)}(T_L)
&=
\frac{N_\sigma(T_L)}
{1+N_\sigma(T_L)}
+\mathcal O(\alpha^4).
\end{align*}
These relations recover the positive-frequency expressions consistently
to the order retained in the weak-drive calculation below.

\subsection{Weak-drive expansion and quadratic rectification law}
\label{appsubsec:single_side_weak_drive}

We start from the sinusoidal flat-spectrum current derived in Sec.~\ref{appsubsec:single_side_flat_spectrum}. In Eq.~\eqref{eq:app_single_side_flat_factorized}, the current is written as the product of the kinetic prefactor $\mathcal K_{\rm flat}$ and the competition bracket $\Xi_{\rm flat}$, whose explicit forms are given in Eqs.~\eqref{eq:app_single_side_Kflat} and \eqref{eq:app_single_side_Xiflat}, respectively. At arbitrary drive strength, the rectification coefficient obtained from these expressions contains the full Floquet sideband sums and is not analytically transparent. The purpose of this appendix is therefore to identify the leading correction to the reciprocal undriven benchmark presented in Sec.~\ref{subsec:undriven_reference_case} and to obtain the coefficient appearing in Eq.~\eqref{eq:single_side_weak_drive_scaling}.

We carry out this expansion around the symmetric static reference point $\omega_{L,0} = \omega_R \equiv\omega_0,\,\gamma_L=\gamma_R$,  for which the sector frequencies reduce to $ \nu_\sigma = \omega_0+\sigma\Omega_J$, \text{and} $\sigma=\pm$.  In the weak-drive regime, $\alpha\ll1$, the expansion proceeds by first expanding the sideband weights in Eq.~\eqref{eq:single_side_sideband_weights} and the effective occupations in Eq.~\eqref{eq:app_single_side_Nsigma}. These determine the ratios in Eq.~\eqref{eq:app_single_side_flat_rsigma}, which enter the competition bracket $\Xi_{\rm flat}$ in Eq.~\eqref{eq:app_single_side_Xiflat}. The positive prefactor $\mathcal K_{\rm flat}$ in Eq.~\eqref{eq:app_single_side_Kflat} is then expanded separately. These ingredients determine the leading correction to the forward and reverse currents, and hence to the rectification coefficient.

For sinusoidal modulation considered in Eq. \eqref{eq:single_side_sinusoidal_drive}, the sideband weights have the weak-drive expansion
\begin{equation}
P(0)
=
1-\frac{\alpha^2}{2}
+
\mathcal O(\alpha^4),
\qquad
P(\pm1)
=
\frac{\alpha^2}{4}
+
\mathcal O(\alpha^4),
\qquad
P(|q|\ge2)
=
\mathcal O(\alpha^4).
\label{eq:app_single_side_Pq_expansion}
\end{equation}
Thus, up to order $\alpha^2$, only the central sideband and the first sidebands contribute. To keep the expansion compact, we define the thermal occupations sampled by these three sidebands as
\begin{equation*}
\bar n_{\sigma,\mu}^{0}
=
\frac{1}{e^{\beta_\mu\hbar\nu_\sigma}-1},
\qquad
\bar n_{\sigma,\mu}^{\pm}
=
\frac{1}{e^{\beta_\mu\hbar(\nu_\sigma\pm\Omega)}-1},
\qquad
\mu\in\{h,c\},
\end{equation*}
together with the finite-difference combination
\begin{equation*}
\Delta\bar n_{\sigma,\mu}
=
\bar n_{\sigma,\mu}^{+}
+
\bar n_{\sigma,\mu}^{-}
-
2\bar n_{\sigma,\mu}^{0}.
\end{equation*}
The competition bracket $\Xi_{\rm flat}$ in Eq.~\eqref{eq:app_single_side_Xiflat} depends on the driven contact through the ratios $r_\sigma^{(L)}$. We therefore first expand the effective occupation entering these ratios. Using Eq.~\eqref{eq:app_single_side_Nsigma} and the sideband expansion in Eq.~\eqref{eq:app_single_side_Pq_expansion}, one obtains
\begin{equation}
N_\sigma(T_\mu)
=
\bar n_{\sigma,\mu}^{0}
+
\frac{\alpha^2}{4}
\Delta\bar n_{\sigma,\mu}
+
\mathcal O(\alpha^4).
\label{eq:app_single_side_Nsigma_expansion}
\end{equation}
Substituting Eq.~\eqref{eq:app_single_side_Nsigma_expansion} into Eq. \eqref{eq:app_single_side_flat_rsigma} gives
\begin{equation}
r_\sigma^{(L)}(T_\mu)
=
e^{-\beta_\mu\hbar\nu_\sigma}
+
\alpha^2 Q_{\sigma,\mu}
+
\mathcal O(\alpha^4),
\label{eq:app_single_side_rsigma_weak}
\end{equation}
where
\begin{equation*}
Q_{\sigma,\mu}
=
\frac{
\Delta\bar n_{\sigma,\mu}
}{
4\bigl(\bar n_{\sigma,\mu}^{0}+1\bigr)^2
}.
\end{equation*}
We now insert Eq.~\eqref{eq:app_single_side_rsigma_weak} into the competition bracket $\Xi_{\rm flat}$ in Eq.~\eqref{eq:app_single_side_Xiflat}. This step isolates the part of the current that controls its sign. For forward bias, $(T_L,T_R)=(T_h,T_c)$, and reverse bias, $(T_L,T_R)=(T_c,T_h)$, $\Xi_{\rm flat}$ in Eq.~\eqref{eq:app_single_side_Xiflat} takes the form
\begin{equation}
\begin{aligned}
\Xi_F
&=
\Xi^{(0)}
+
\alpha^2\Xi_F^{(2)}
+
\mathcal O(\alpha^4),
\qquad\qquad
\Xi_B
=
-\Xi^{(0)}
+
\alpha^2\Xi_B^{(2)}
+
\mathcal O(\alpha^4).
\end{aligned}
\label{eq:app_single_side_Xi_FB_expansion}
\end{equation}
Here the zeroth-order contribution is
\begin{equation*}
\Xi^{(0)}
=
2e^{-(\beta_h+\beta_c)\hbar\omega_0}
\sinh\!\big[
(\beta_h-\beta_c)\hbar\Omega_J
\big],
\end{equation*}
and the leading drive-induced corrections are
\begin{equation*}
\begin{aligned}
\Xi_F^{(2)}
&=
e^{-\beta_c\hbar\nu_+}Q_{-,h}
-
e^{-\beta_c\hbar\nu_-}Q_{+,h},
\qquad\qquad
\Xi_B^{(2)}
=
e^{-\beta_h\hbar\nu_+}Q_{-,c}
-
e^{-\beta_h\hbar\nu_-}Q_{+,c}.
\end{aligned}
\end{equation*}
The opposite signs of the zeroth-order terms in Eq.~\eqref{eq:app_single_side_Xi_FB_expansion} reproduce the reciprocal undriven benchmark: before the drive-induced corrections are included, reversing the thermal bias changes only the sign of the current.
The kinetic prefactor in Eq.~\eqref{eq:app_single_side_Kflat} also receives a weak-drive correction. Since this prefactor is positive, it does not determine the blocking condition, but its forward--reverse asymmetry contributes to the magnitude of the rectification coefficient. We therefore write
\begin{equation}
\mathcal K_{\rm flat}(T_L,T_R)
=
\mathcal K^{(0)}
\bigl[
1+\alpha^2\chi_{\mathcal K}(T_L,T_R)
\bigr]
+
\mathcal O(\alpha^4),
\label{eq:app_single_side_Kexp}
\end{equation}
where $\mathcal K^{(0)}$ is the common zeroth-order value for the forward and reverse biases at the symmetric static reference point. From Eq.~\eqref{eq:app_single_side_Kflat}, the relative correction is
\begin{equation*}
\chi_{\mathcal K}(T_L,T_R)
=
\frac{\Delta\bar n_{+,L}}{4(\bar n_{+,L}^{0}+1)}
+
\frac{\Delta\bar n_{-,L}}{4(\bar n_{-,L}^{0}+1)}
-
\frac{\mathcal Z_{\mathrm{sd}}^{(2)}(T_L,T_R)}
{\mathcal Z_{\mathrm{sd}}^{(0)}(T_L,T_R)} .
\end{equation*}
Here $\bar n_{\sigma,L}^{0}$ and $\Delta\bar n_{\sigma,L}$ are evaluated at the left-bath temperature, and the normalization factor has been expanded as
\begin{equation*}
\mathcal Z_{\mathrm{sd}}(T_L,T_R)
=
\mathcal Z_{\mathrm{sd}}^{(0)}(T_L,T_R)
+
\alpha^2
\mathcal Z_{\mathrm{sd}}^{(2)}(T_L,T_R)
+
\mathcal O(\alpha^4).
\end{equation*}
Combining the competition-bracket expansion in Eq.~\eqref{eq:app_single_side_Xi_FB_expansion} with the kinetic-prefactor expansion in Eq.~\eqref{eq:app_single_side_Kexp} and inserting this into Eq. \eqref{eq:app_single_side_flat_factorized}, the forward and reverse currents become
\begin{equation}
\begin{aligned}
\mathcal J_F
&=
2\hbar\Omega_J\,\mathcal K^{(0)}
\Big[
\Xi^{(0)}
+
\alpha^2
\bigl(
\Xi_F^{(2)}
+
\chi_{\mathcal K,F}\Xi^{(0)}
\bigr)
\Big]
+
\mathcal O(\alpha^4),
\\
\mathcal J_B
&=
2\hbar\Omega_J\,\mathcal K^{(0)}
\Big[
-\Xi^{(0)}
+
\alpha^2
\bigl(
\Xi_B^{(2)}
-
\chi_{\mathcal K,B}\Xi^{(0)}
\bigr)
\Big]
+
\mathcal O(\alpha^4).
\end{aligned}
\label{eq:app_single_side_JFB_weak}
\end{equation}
Here, we denote the forward- and reverse-bias corrections by $\chi_{\mathcal K,F}=\chi_{\mathcal K}(T_h,T_c)$, and $\chi_{\mathcal K,B}=\chi_{\mathcal K}(T_c,T_h)$, respectively.
At zeroth order, these currents have equal magnitude and opposite sign, reproducing the reciprocal undriven benchmark presented in Sec. \ref{subsec:undriven_reference_case}. Hence, the numerator of the rectification coefficient \eqref{eq:results_rectification_coefficient}, $|\mathcal J_F+\mathcal J_B|$, receives its first nonzero contribution at order $\alpha^2$, while the denominator remains finite at zeroth order. Substituting Eq.~\eqref{eq:app_single_side_JFB_weak} into Eq.~\eqref{eq:results_rectification_coefficient} gives
\begin{equation}
\mathcal R
=
\alpha^2
\mathcal F(T_h,T_c,\omega_0,J,\Omega)
+
\mathcal O(\alpha^4),
\label{eq:app_single_side_Rmain}
\end{equation}
where
\begin{equation}
\mathcal F(T_h,T_c,\omega_0,J,\Omega)
=
\left|
\frac{
\Xi_F^{(2)}
+
\Xi_B^{(2)}
}{
\Xi^{(0)}
}
+
\chi_{\mathcal K,F}-\chi_{\mathcal K,B}
\right|.
\label{eq:app_single_side_Fcoef}
\end{equation}

\section{Dual-side-driven steady-state current}
\label{appsubsec:dual_side_current}

This appendix derives the steady-state heat current entering the right reservoir in the dual-side-driven configuration presented in Sec. \ref{subsec:dual_side_driving}. We start from the general Floquet heat-current definition in Eq.~\eqref{eq:results_heat_current_definition} and rewrite it in the four-state population basis, where the relevant transport cycle and the sideband-resolved right-contact transitions can be identified explicitly. The resulting expression is then reduced to a rate-only form and expanded in the weak-drive flat-spectrum limit to obtain the dual-side-driven current formula.

\subsection{Heat-current derivation}
\label{appsubsec:dual_side_heat_current_derivation}

We first evaluate the right-reservoir heat current in the four-state population basis. The basis labels \(A=|00\rangle\), \(B=|10\rangle\), \(C=|11\rangle\), and \(D=|01\rangle\) allow us to identify the transition graph and the loop that connects the two reservoirs. With this convention, the population loop relevant for transferring energy between the reservoirs is
\begin{equation}
A
\xrightarrow{\,u_L^-\,}
B
\xrightarrow{\,u_R^+\,}
C
\xrightarrow{\,d_L^+\,}
D
\xrightarrow{\,d_R^-\,}
A .
\label{eq:app_dual_transport_cycle}
\end{equation}
The labels above the arrows denote the total rates defined in Eq.~\eqref{eq:dual_side_effective_rates}. Each total rate is obtained by
summing the corresponding sideband-resolved rate over an independent Floquet index. The cycle in Eq.~\eqref{eq:app_dual_transport_cycle} shows that heat transfer between the reservoirs requires a closed population loop involving both left- and right-contact transitions. By contrast, a local back-and-forth process such as \(A\rightarrow B\rightarrow A\) does not transfer heat between the reservoirs. It excites and de-excites the same transition through the same contact, and therefore does not complete a loop connecting the two reservoirs.

Having identified the transport loop, we now evaluate the energy exchanged along its right-contact edges. The population-basis current follows by applying the Floquet Hamiltonian in Eq.~\eqref{eq:results_floquet_hamiltonian} to the right-contact transitions. A right-bath transition in sector \(\sigma=\pm\) and sideband \(q\) exchanges the energy \(\hbar(\nu_R^{(\sigma)}+q\Omega)\) with the right reservoir. Using the heat-current definition in Eq.~\eqref{eq:results_heat_current_definition}, each sideband-resolved net probability flow is therefore weighted by the corresponding exchanged energy:
\begin{align}
\mathcal{J}_R^{\rm ss}
=&\,
\hbar
\sum_{p\in\mathbb Z}
\left(\nu_R^{(+)}+p\Omega\right)
\left[
u_{R,p}^{+}p_B^{\rm ss}
-
d_{R,p}^{+}p_C^{\rm ss}
\right]
\nonumber\\
&+
\hbar
\sum_{p\in\mathbb Z}
\left(\nu_R^{(-)}+p\Omega\right)
\left[
u_{R,p}^{-}p_A^{\rm ss}
-
d_{R,p}^{-}p_D^{\rm ss}
\right].
\label{eq:app_dual_expanded_right_current}
\end{align}
The two terms correspond to the two conditional sectors of the right-qubit transition. The first term comes from the \(B\leftrightarrow C\) edge, where the right qubit changes state while the left qubit is excited. The second term comes from the \(A\leftrightarrow D\) edge, where the same right-qubit transition occurs while the left qubit is in its ground state. Here \(p_X^{\rm ss}\) denotes the steady-state population of state \(X=A,B,C,D\).

The two factors \(\nu_R^{(\sigma)}\) and \(q\Omega\) in Eq.~\eqref{eq:app_dual_expanded_right_current} have different physical origins. The first is the conditional transition energy, while the second is the Floquet sideband energy. This separation allows the current to be written as the sum of a transition-energy contribution and a sideband-energy contribution:
\begin{equation}
\mathcal{J}_R^{\rm ss}
=
\mathcal{J}_{R,\rm tr}^{\rm ss}
+
\mathcal{J}_{R,\rm pump}^{\rm ss}.
\label{eq:app_dual_current_split}
\end{equation}
The first term, \(\mathcal{J}_{R,\rm tr}^{\rm ss}\), is the transition-energy contribution. The second term, \(\mathcal{J}_{R,\rm pump}^{\rm ss}\), is the sideband-energy contribution generated by the driven right contact. To express the transition-energy part in terms of the stationary population dynamics, we introduce the total right-contact rates
\begin{equation*}
u_R^\sigma=\sum_q u_{R,q}^{(\sigma)},
\qquad
d_R^\sigma=\sum_q d_{R,q}^{(\sigma)} .
\end{equation*}
Using these rates, the transition-energy contribution is
\begin{align}
\mathcal{J}_{R,\rm tr}^{\rm ss}
=&\,
\hbar\nu_R^{(+)}
\left[
u_R^+p_B^{\rm ss}
-
d_R^+p_C^{\rm ss}
\right]
+
\hbar\nu_R^{(-)}
\left[
u_R^-p_A^{\rm ss}
-
d_R^-p_D^{\rm ss}
\right].
\label{eq:app_dual_transition_current}
\end{align}
The part of Eq.~\eqref{eq:app_dual_expanded_right_current} proportional to \(q\Omega\) gives the sideband-energy contribution:
\begin{align}
\mathcal{J}_{R,\rm pump}^{\rm ss}
=&\,
\hbar\Omega
\sum_q q
\left[
u_{R,q}^{(+)}p_B^{\rm ss}
-
d_{R,q}^{(+)}p_C^{\rm ss}
\right]
+
\hbar\Omega
\sum_q q
\left[
u_{R,q}^{(-)}p_A^{\rm ss}
-
d_{R,q}^{(-)}p_D^{\rm ss}
\right].
\label{eq:app_dual_pump_current_sum}
\end{align}
We now simplify these two contributions separately: the transition-energy part reduces to a cycle current, whereas the sideband-energy part reduces to a pumping term.

We first simplify the transition-energy part. Since Eq.~\eqref{eq:app_dual_transition_current} contains the net probability flows through the two right-contact edges, it is useful to express these flows in terms of the stationary cycle current. For the positive orientation \(A\rightarrow B\rightarrow C\rightarrow D\rightarrow A\), the directed edge currents are
\begin{align*}
I_{AB}
=
u_L^-p_A^{\rm ss}
-
d_L^-p_B^{\rm ss}, \qquad\quad
I_{BC}
=
u_R^+p_B^{\rm ss}
-
d_R^+p_C^{\rm ss}, \qquad\quad
I_{CD}
=
d_L^+p_C^{\rm ss}
-
u_L^+p_D^{\rm ss}, \qquad\quad
I_{DA}=
d_R^-p_D^{\rm ss}
-
u_R^-p_A^{\rm ss}.
\end{align*}
Each edge current is the probability flow along the chosen cycle direction minus the flow in the opposite direction. At steady state, no probability accumulates at any vertex of the four-state graph. Therefore, the same net probability current must pass through every edge of the closed loop:
\begin{equation}
I_{AB}=I_{BC}=I_{CD}=I_{DA}\equiv I_{\rm cyc}.
\label{eq:app_dual_cycle_current_definition}
\end{equation}
This equality is a consequence of stationarity, not of detailed balance. With the chosen orientation, the \(B\leftrightarrow C\) right-contact edge contributes \(+I_{\rm cyc}\) to Eq.~\eqref{eq:app_dual_transition_current}, whereas the \(A\leftrightarrow D\) edge contributes \(-I_{\rm cyc}\). Substituting Eq.~\eqref{eq:app_dual_cycle_current_definition} into Eq.~\eqref{eq:app_dual_transition_current} and using \(\nu_R^{(+)}-\nu_R^{(-)}=2\Omega_J\) gives
\begin{equation}
\mathcal{J}_{R,\rm tr}^{\rm ss}
=
2\hbar\Omega_J I_{\rm cyc}.
\label{eq:app_dual_transition_current_final}
\end{equation}
Thus, the transition-energy contribution is controlled by the interaction-induced splitting \(2\Omega_J\) and the stationary cycle current. We next simplify the sideband-energy part in Eq.~\eqref{eq:app_dual_current_split}. The \(q\Omega\)-weighted terms are naturally collected by the first sideband moments of the right-contact rates, defined as
\begin{equation}
\widetilde u_R^\sigma
=
\sum_{p\in\mathbb Z}p\,u_{R,p}^{\sigma},
\qquad
\widetilde d_R^\sigma
=
\sum_{p\in\mathbb Z}p\,d_{R,p}^{\sigma}.
\label{eq:app_dual_sideband_moments}
\end{equation}
These moments measure the upward and downward transition rates weighted by the Floquet sideband index \(q\). They therefore collect the part of the current associated with the sideband energy \(q\Omega\). Using Eq.~\eqref{eq:app_dual_sideband_moments}, the sideband-energy current in Eq.~\eqref{eq:app_dual_pump_current_sum} can be written as
\begin{equation}
\mathcal{J}_{R,\rm pump}^{\rm ss}
=
\hbar\Omega \mathcal P_R,
\label{eq:app_dual_pump_current_final}
\end{equation}
where
\begin{equation*}
\mathcal P_R
=
\widetilde u_R^+p_B^{\rm ss}
-
\widetilde d_R^+p_C^{\rm ss}
+
\widetilde u_R^-p_A^{\rm ss}
-
\widetilde d_R^-p_D^{\rm ss}.
\end{equation*}
\(\mathcal P_R\) is the net sideband-weighted probability flow through the right-contact transitions. It vanishes when the right contact is not driven, because only the \(q=0\) sideband contributes, and all first-sideband moments are zero. Combining Eqs.~\eqref{eq:app_dual_transition_current_final} and \eqref{eq:app_dual_pump_current_final}, the right-reservoir current becomes
\begin{equation}
\mathcal{J}_R^{\rm ss}
=
2\hbar\Omega_J I_{\rm cyc}
+
\hbar\Omega \mathcal P_R .
\label{eq:app_dual_cycle_plus_pump_current}
\end{equation}
Equation~\eqref{eq:app_dual_cycle_plus_pump_current} separates the current into an interaction-mediated cycle contribution and a right-contact Floquet-pumping contribution. The first term is present when the stationary population loop carries a nonzero current. The second term is absent in the single-side-driven configuration considered in Sec.~\ref{subsec:single_side_driving}, but generally survives when the right contact is also modulated.

Solving the stationary four-state rate problem expresses both \(I_{\rm cyc}\) and \(\mathcal P_R\) in terms of transition rates. Substitution into Eq.~\eqref{eq:app_dual_cycle_plus_pump_current} gives the compact rate-only form
\begin{equation}
\mathcal{J}_R^{\rm ss}
=
\frac{\hbar}{Z_{\rm dd}}
\left[
2\Omega_J
\left(\mathcal{C}_{\rm f}
-
\mathcal{C}_{\rm b}
\right)
+
\Omega\mathcal B_R
\right].
\label{eq:app_dual_final_right_current}
\end{equation}
The quantities appearing in Eq.~\eqref{eq:app_dual_final_right_current} are defined as follows. The stationary cycle current is
\begin{align}
I_{\rm cyc}
=
\frac{\mathcal{C}_{\rm f}-\mathcal{C}_{\rm b}}{Z_{\rm dd}}, \quad\text{and}\quad
\mathcal{C}_{\rm f}
=
u_L^-u_R^+d_L^+d_R^-,
\qquad
\mathcal{C}_{\rm b}
=
u_L^+u_R^-d_L^-d_R^+ .
\label{eq:app_dual_cycle_current_rate_form}
\end{align}
The quantities \(\mathcal{C}_{\rm f}\) and \(\mathcal{C}_{\rm b}\) are the products of rates around the forward and backward orientations of the population loop, respectively. The positive denominator \(Z_{\rm dd}\) is the normalization factor of the stationary four-state problem and has the form
\begin{equation}
Z_{\rm dd}=Q_A+Q_B+Q_C+Q_D .
\label{eq:app_dual_Zdd_definition}
\end{equation}
Here \(Q_X\) are the unnormalized stationary populations:
\begin{align}
Q_A
&=
d_L^- d_R^+ u_L^+
+
d_L^- d_R^+ d_R^-
+
d_L^- d_L^+ d_R^-
+
u_R^+ d_L^+ d_R^-,
\qquad
Q_B
=
u_L^- d_R^+ u_L^+
+
u_L^- d_R^+ d_R^-
+
u_L^- d_L^+ d_R^-
+
d_R^+ u_L^+ u_R^-,
\nonumber\\
Q_C
&=
u_L^- u_R^+ u_L^+
+
u_L^- u_R^+ d_R^-
+
d_L^- u_L^+ u_R^-
+
u_R^+ u_L^+ u_R^-,
\qquad
Q_D
=
u_L^- u_R^+ d_L^+
+
d_L^- d_R^+ u_R^-
+
d_L^- d_L^+ u_R^-
+
u_R^+ d_L^+ u_R^- .
\label{eq:app_dual_Q_definitions}
\end{align}
The sideband-weighted pumping numerator appearing in Eq.~\eqref{eq:app_dual_final_right_current} is
\begin{equation}
\mathcal B_R
=
\widetilde u_R^+Q_B
-
\widetilde d_R^+Q_C
+
\widetilde u_R^-Q_A
-
\widetilde d_R^-Q_D .
\label{eq:app_dual_BR_definition}
\end{equation}
Equation~\eqref{eq:app_dual_final_right_current} is therefore the rate-only form of the dual-side-driven right-reservoir current. Its first term is governed by the imbalance between the forward and backward population cycles, whereas its second term is governed by the sideband-weighted pumping numerator at the driven right contact. Consequently, canceling the cycle imbalance, \(\mathcal{C}_{\rm f}=\mathcal{C}_{\rm b}\), does not by itself guarantee zero current unless the pumping contribution is also canceled.

\subsection{Weak-drive expansion of the dual-side-driven current}
\label{app:dual_side_weak_drive_expansion}

The purpose of this subsection is to derive the weak-drive expansion of the exact right-bath heat current for the dual-side-driven configuration. We take the exact right-bath current in Eq.~\eqref{eq:app_dual_final_right_current} as the starting point. Throughout this subsection, we use the symmetric resonant reference point \(\omega_{L,0}=\omega_{R,0}\equiv\omega_0\) and \(\kappa_L=\kappa_R\equiv\kappa\), where \(\kappa_a\) is the effective flat-spectrum transition-rate prefactor of the bath given in Eq. \eqref{eq:app_single_side_kappas}. We also assume weak modulation, \(\alpha_a\ll1\). Using the weak-drive sideband weights in Eq.~\eqref{eq:dual_side_weak_sideband_weights}, only the carrier and the first sidebands contribute through second order in \(\alpha_a\).

Expanding the exact total rates in Eq.~\eqref{eq:dual_side_effective_rates} with these weak sideband weights gives
\begin{align}
u_a^\sigma
&=
\kappa
\left[
n_{a,\sigma}
+
\alpha_a^2D_{a,\sigma}
\right]
+
\mathcal O(\alpha_a^4),
\nonumber\\
d_a^\sigma
&=
\kappa
\left[
1+n_{a,\sigma}
+
\alpha_a^2D_{a,\sigma}
\right]
+
\mathcal O(\alpha_a^4).
\label{eq:app_dual_weak_rates}
\end{align}
Here
\begin{equation*}
n_{a,\sigma}
=
\bar n_a(\nu_\sigma),
\qquad
n_{a,\sigma}^{\pm}
=
\bar n_a(\nu_\sigma\pm\Omega),
\qquad
D_{a,\sigma}
=
\frac{1}{4}
\left(
n_{a,\sigma}^{+}
+
n_{a,\sigma}^{-}
-
2n_{a,\sigma}
\right).
\end{equation*}
The exact right-bath current in Eq.~\eqref{eq:app_dual_final_right_current} also contains terms proportional to the sideband energy \(q\Omega\). Therefore, in addition to the total rates, we need the first sideband moments defined in Eq.~\eqref{eq:app_dual_sideband_moments}. Their weak-drive form is
\begin{equation}
\widetilde u_R^\sigma
=
\widetilde d_R^\sigma
=
\kappa\alpha_R^2M_{R,\sigma}
+
\mathcal O(\alpha_R^4),
\qquad
M_{R,\sigma}
=
\frac{1}{4}
\left(
n_{R,\sigma}^{+}
-
n_{R,\sigma}^{-}
\right).
\label{eq:app_dual_weak_moment_rates}
\end{equation}
We now insert Eq.~\eqref{eq:app_dual_weak_rates} into the forward and backward cycle products defined in Eq.~\eqref{eq:app_dual_cycle_current_rate_form}. Their difference expands as
\begin{equation}
\mathcal{C}_{\rm f}
-
\mathcal{C}_{\rm b}
=
\kappa^4
\left[
\delta_0
+
\alpha_L^2\delta_L
+
\alpha_R^2\delta_R
+
\mathcal O
\left(
\alpha_L^4,
\alpha_R^4,
\alpha_L^2\alpha_R^2
\right)
\right].
\label{eq:app_dual_cycle_imbalance_weak}
\end{equation}
The zeroth-order imbalance is
\begin{equation*}
\delta_0
=
\mathsf a\mathsf b\mathsf c\mathsf d
-
\bar{\mathsf a}\bar{\mathsf b}\bar{\mathsf c}\bar{\mathsf d},
\end{equation*}
while the left- and right-drive corrections are
\begin{align*}
\delta_L
&=
D_{L,-}
\left(
\mathsf b\mathsf c\mathsf d
-
\bar{\mathsf b}\bar{\mathsf c}\bar{\mathsf d}
\right)
+
D_{L,+}
\left(
\mathsf a\mathsf b\mathsf d
-
\bar{\mathsf a}\bar{\mathsf b}\bar{\mathsf d}
\right),
\qquad
\delta_R
=
D_{R,+}
\left(
\mathsf a\mathsf c\mathsf d
-
\bar{\mathsf a}\bar{\mathsf c}\bar{\mathsf d}
\right)
+
D_{R,-}
\left(
\mathsf a\mathsf b\mathsf c
-
\bar{\mathsf a}\bar{\mathsf b}\bar{\mathsf c}
\right).
\end{align*}
The compact variables denote the zeroth-order dimensionless rates along the forward and reverse cycles:
\begin{equation*}
\mathsf a=n_{L,-},
\quad
\bar{\mathsf a}=1+n_{L,-},
\qquad
\mathsf b=n_{R,+},
\quad
\bar{\mathsf b}=1+n_{R,+},
\qquad
\mathsf c=1+n_{L,+},
\quad
\bar{\mathsf c}=n_{L,+},
\qquad
\mathsf d=1+n_{R,-},
\quad
\bar{\mathsf d}=n_{R,-}.
\end{equation*}
Because the current~\eqref{eq:app_dual_final_right_current} is a ratio, the normalization denominator in Eq.~\eqref{eq:app_dual_Zdd_definition} must be expanded to the same order. We write
\begin{equation}
Z_{\rm dd}
=
\kappa^3
\left[
z_0
+
\alpha_L^2z_L
+
\alpha_R^2z_R
+
\mathcal O
\left(
\alpha_L^4,
\alpha_R^4,
\alpha_L^2\alpha_R^2
\right)
\right].
\label{eq:app_dual_Z_weak_expansion}
\end{equation}
The zeroth-order denominator is
\begin{equation*}
z_0
=
\mathcal Z
\left(
\mathsf a,\mathsf b,\mathsf c,\mathsf d;
\bar{\mathsf a},\bar{\mathsf b},\bar{\mathsf c},\bar{\mathsf d}
\right),
\end{equation*}
where \(\mathcal Z\) is the cubic normalization polynomial obtained by summing the four zeroth-order unnormalized populations:
\begin{align*}
\mathcal Z
={}&
x_ax_bx_c
+
x_ax_b\bar x_c
+
x_ax_bx_d
+
x_a\bar x_b\bar x_c
+
x_a\bar x_bx_d
+
x_ax_cx_d
+
\bar x_a\bar x_b\bar x_c
+
\bar x_a\bar x_bx_d
\nonumber\\
&+
\bar x_a\bar x_b\bar x_d
+
\bar x_ax_cx_d
+
\bar x_ax_c\bar x_d
+
\bar x_a\bar x_c\bar x_d
+
x_bx_cx_d
+
x_bx_c\bar x_d
+
x_b\bar x_c\bar x_d
+
\bar x_b\bar x_c\bar x_d .
\end{align*}
The denominator corrections are obtained by Taylor expanding \(\mathcal Z\) under the weak-drive shifts of the rate variables. The left-drive correction is
\begin{equation*}
z_L
=
D_{L,-}
\left[
\left(
\frac{\partial\mathcal Z}{\partial x_a}
\right)_0
+
\left(
\frac{\partial\mathcal Z}{\partial \bar x_a}
\right)_0
\right]
+
D_{L,+}
\left[
\left(
\frac{\partial\mathcal Z}{\partial x_c}
\right)_0
+
\left(
\frac{\partial\mathcal Z}{\partial \bar x_c}
\right)_0
\right],
\end{equation*}
and the right-drive correction is
\begin{equation*}
z_R
=
D_{R,+}
\left[
\left(
\frac{\partial\mathcal Z}{\partial x_b}
\right)_0
+
\left(
\frac{\partial\mathcal Z}{\partial \bar x_b}
\right)_0
\right]
+
D_{R,-}
\left[
\left(
\frac{\partial\mathcal Z}{\partial x_d}
\right)_0
+
\left(
\frac{\partial\mathcal Z}{\partial \bar x_d}
\right)_0
\right].
\end{equation*}
The subscript \(0\) means that the derivatives are evaluated at
\begin{equation*}
(x_a,x_b,x_c,x_d;\bar x_a,\bar x_b,\bar x_c,\bar x_d)
=
(\mathsf a,\mathsf b,\mathsf c,\mathsf d;
\bar{\mathsf a},\bar{\mathsf b},\bar{\mathsf c},\bar{\mathsf d}).
\end{equation*}
It remains to expand the right-contact pumping numerator defined in Eq.~\eqref{eq:app_dual_BR_definition}. Since the sideband moments in Eq.~\eqref{eq:app_dual_weak_moment_rates} are already of order \(\alpha_R^2\), the unnormalized populations entering this term are needed only at zeroth order. Hence,
\begin{equation}
\mathcal B_R
=
\alpha_R^2\kappa^4 b_R
+
\mathcal O
\left(
\alpha_R^4,
\alpha_L^2\alpha_R^2
\right).
\label{eq:app_dual_BR_weak}
\end{equation}
To define \(b_R\), we write the zeroth-order unnormalized populations as \(Q_X^{(0)}=\kappa^3q_X\). From Eq.~\eqref{eq:app_dual_Q_definitions}, their dimensionless forms are
\begin{align*}
q_A
&=
\bar{\mathsf a}\bar{\mathsf b}\bar{\mathsf c}
+
\bar{\mathsf a}\bar{\mathsf b}\mathsf d
+
\bar{\mathsf a}\mathsf c\mathsf d
+
\mathsf b\mathsf c\mathsf d,
\qquad
q_B
=
\mathsf a\bar{\mathsf b}\bar{\mathsf c}
+
\mathsf a\bar{\mathsf b}\mathsf d
+
\mathsf a\mathsf c\mathsf d
+
\bar{\mathsf b}\bar{\mathsf c}\bar{\mathsf d},
\nonumber\\
q_C
&=
\mathsf a\mathsf b\bar{\mathsf c}
+
\mathsf a\mathsf b\mathsf d
+
\bar{\mathsf a}\bar{\mathsf c}\bar{\mathsf d}
+
\mathsf b\bar{\mathsf c}\bar{\mathsf d},
\qquad
q_D
=
\mathsf a\mathsf b\mathsf c
+
\bar{\mathsf a}\bar{\mathsf b}\bar{\mathsf d}
+
\bar{\mathsf a}\mathsf c\bar{\mathsf d}
+
\mathsf b\mathsf c\bar{\mathsf d}.
\end{align*}
The pumping coefficient is then
\begin{equation*}
b_R
=
M_{R,+}(q_B-q_C)
+
M_{R,-}(q_A-q_D).
\end{equation*}
Finally, inserting Eqs.~\eqref{eq:app_dual_cycle_imbalance_weak}, \eqref{eq:app_dual_Z_weak_expansion}, and \eqref{eq:app_dual_BR_weak} into Eq.~\eqref{eq:app_dual_final_right_current}, and expanding the denominator consistently to second order, gives
\begin{equation}
\mathcal J_{R,\rm flat}^{\rm ss}
=
\mathcal J_0
+
\alpha_L^2\mathcal J_{\mathrm L}^{(2)}
+
\alpha_R^2\mathcal J_{\mathrm R}^{(2)}
+
\mathcal O
\left(
\alpha_L^4,
\alpha_R^4,
\alpha_L^2\alpha_R^2
\right).
\label{eq:app_dual_weak_current_final}
\end{equation}
The zeroth-order term is
\begin{equation*}
\mathcal J_0
=
2\hbar\Omega_J\kappa
\frac{\delta_0}{z_0},
\end{equation*}
and the quadratic corrections are
\begin{align}
\mathcal J_{\mathrm L}^{(2)}
&=
2\hbar\Omega_J\kappa
\left[
\frac{\delta_L}{z_0}
-
\frac{\delta_0z_L}{z_0^2}
\right],
\qquad
\mathcal J_{\mathrm R}^{(2)}
=
\hbar\kappa
\left[
2\Omega_J
\left(
\frac{\delta_R}{z_0}
-
\frac{\delta_0z_R}{z_0^2}
\right)
+
\Omega
\frac{b_R}{z_0}
\right].
\label{eq:app_dual_quadratic_corrections}
\end{align}

\end{widetext}

\bibliography{FQTD}

\end{document}